%% file: paper.tex
\documentclass[aps,prb,amsmath,amssymb,
superscriptaddress,floatfix,showpacs,twocolumn]{revtex4-1}

\newcommand{\linecite}[1]{Ref.~\onlinecite{#1}}

\newcommand{\ee}[1]{  \textrm{e}^{#1 } }

\newcommand{\bb}{\mathbf}

\newcommand{\xvec}{\hat{\bf x}}
\newcommand{\yvec}{\hat{\bf y}}

\usepackage{graphicx}
\usepackage{empheq}
\usepackage{ amssymb}
\usepackage{ amsmath}
\usepackage{color}
\usepackage{dsfont}
\usepackage[utf8]{inputenc}
\usepackage[T1]{fontenc}
\usepackage[ocgcolorlinks, linkcolor=blue,citecolor=blue]{hyperref}

\begin{document}

\title{Novel phases in a square--lattice frustrated ferromagnet : $1/3$--magnetisation plateau,
	helicoidal spin--liquid and vortex crystal}

\author{Luis Seabra}

\affiliation{Department of Physics, Technion--Israel Institute of Technology, Haifa 32000, Israel}

\affiliation{H.\ H.\ Wills Physics Laboratory, University of Bristol,  Tyndall Av, BS8--1TL, UK}

\author{Philippe Sindzingre}

\affiliation{Laboratoire de Physique Th\'eorique de la Mati\`ere Condens\'ee,
UMR 7600, CNRS, UPMC Univ. Paris 06,  Sorbonne Universit\'es, case 121,
4 Place Jussieu, FR--75252 Paris Cedex, France}

\author{Tsutomu Momoi}

\affiliation{Condensed Matter Theory Laboratory, RIKEN, Wako, Saitama 351-0198, Japan}

\affiliation{RIKEN Center for Emergent Matter Science (CEMS), Wako, Saitama 351-0198, Japan}

\author{Nic Shannon}

\affiliation{H.\ H.\ Wills Physics Laboratory, University of Bristol,  Tyndall Av, BS8--1TL, UK}

\affiliation{Okinawa Institute for Science and Technology Graduate University, Onna-son,
Okinawa 904-0495, Japan}

\date{\today}

\begin{abstract}
A large part of the interest in magnets with frustrated antiferromagnetic interactions
comes from the many new phases found in applied magnetic field.
In this Article, we explore some of the new  phases which arise
in a model with frustrated {\it ferromagnetic} interactions, the \mbox{$J_1$--$J_2$}--$J_3$
Heisenberg model on a square lattice.
Using a combination of classical Monte--Carlo simulation and spin--wave theory,
we uncover behaviour reminiscent of some
widely--studied frustrated antiferromagnets, but
with a number of new twists.
We first demonstrate that, for a suitable choice of parameters,
the phase diagram as a function of magnetic field and temperature
is nearly identical to that of the Heisenberg antiferromagnet
on a triangular lattice, including the celebrated \mbox{$1/3$--magnetisation} plateau.
We then examine how this phase diagram changes when the model is tuned to a point
where the classical ground--state is highly degenerate.
In this case, two new phases emerge; a classical, finite--temperature spin--liquid,
characterised by a ``ring'' in the spin structure--factor ${\mathcal S}({\bf q})$; and a vortex crystal,
a multiple--Q state with finite magnetisation, which can be viewed as an ordered
lattice of magnetic vortices.
All of these new phases persist for a wide range of magnetic field.
We discuss the relationship between these results and published studies
of frustrated antiferromagnets, together with some of the materials where these new phases
might be observed in experiment.
\end{abstract}

\pacs{75.10.Hk, 67.80.kb,  74.25.Uv}

\maketitle

\section{Introduction}
\label{section:introduction}


Much attention has been devoted to the question of whether a frustrated magnet
orders or not~\cite{Balents2010}.
Even in the cases where such systems do order, the results are
often surprising and unconventional.
Novel phases of matter have often been uncovered by applying an external magnetic field
to frustrated magnet systems~\cite{Moessner2009,Starykh2014}.
Examples of these phases range from quasi-classical magnetisation plateaux and  magnetic
supersolid phases~\cite{Kawamura1985,Matsuda1970,Liu1973,Seabra2010},  magnetic
analogues of liquid crystals~\cite{Andreev1984,Shannon2006,Smerald2013},
and crystals formed of skyrmions, which have already been observed in
experiment~\cite{Muhlbauer2009,Schulz2012}.


This ongoing interest in novel magnetic phases is underwritten by a
steady stream of new materials.
Among these, materials with frustrated ferromagnetic interactions, i.e. with
ferromagnetic largest interactions but with a
non-ferromagnetic ground state driven by competing interactions,
have a particularly rich phenomenology.
Well--known examples include spin ice, a celebrated example of a
three--dimensional classical spin liquid
 ~\cite{castelnovo12}, and the solid phases
of $^3$He, whose nuclear magnetism continues to push the limits of
our understanding of quantum spins~\cite{momoi12,casey13}.
More recent discoveries include new families of layered, square--lattice
vanadates~\cite{Kaul2004,Nath2008,Skoulatos2009}
and cuprates~\cite{Kageyama2005,Tsujimoto2007,tassel10,Yusuf2011},
with properties that encompass
ordered ground states selected by fluctuations~\cite{Skoulatos2009},
exotic singlet phases~\cite{tassel10},
helical order~\cite{Yusuf2011},
and surprisingly, a $1/3$--magnetisation plateau~\cite{Tsujimoto2007}.

\begin{figure*}[t!]
\centering
\includegraphics[width=0.8\paperwidth]{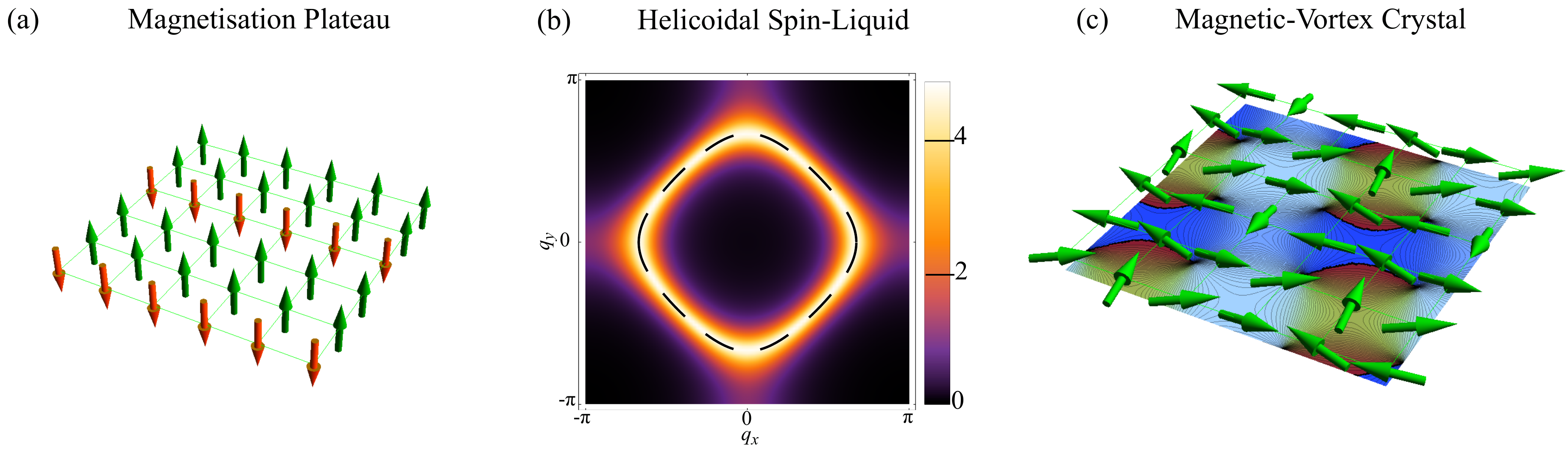}
\caption{
(Color online).
Novel phases found in a square--lattice frustrated ferromagnet  in applied magnetic field.
(a) Collinear $1/3$--magnetisation plateau, with stripe--like three--sublattice order.
(b) Spin liquid with short-range helicoidal correlations, characterised by a ``ring'' in the spin
     structure factor ${\mathcal S}({\bf q})$, shown here for \mbox{$h=0$}.
(c) Crystal of magnetic vortices, with finite magnetisation.
The colour map is a representation of the phase of the spin texture in the $S^x$--$S^y$ plane.
The model studied is the \mbox{$J_1$--$J_2$}--$J_3$  Heisenberg model on a square lattice,
$\mathcal{H}^{\sf FFM}_\square$~[Eq.~(\ref{eq:Hex})].
}
\label{fig:novel.phases}
\end{figure*}
\begin{figure}[t!]
\centering
\includegraphics[width=\columnwidth]{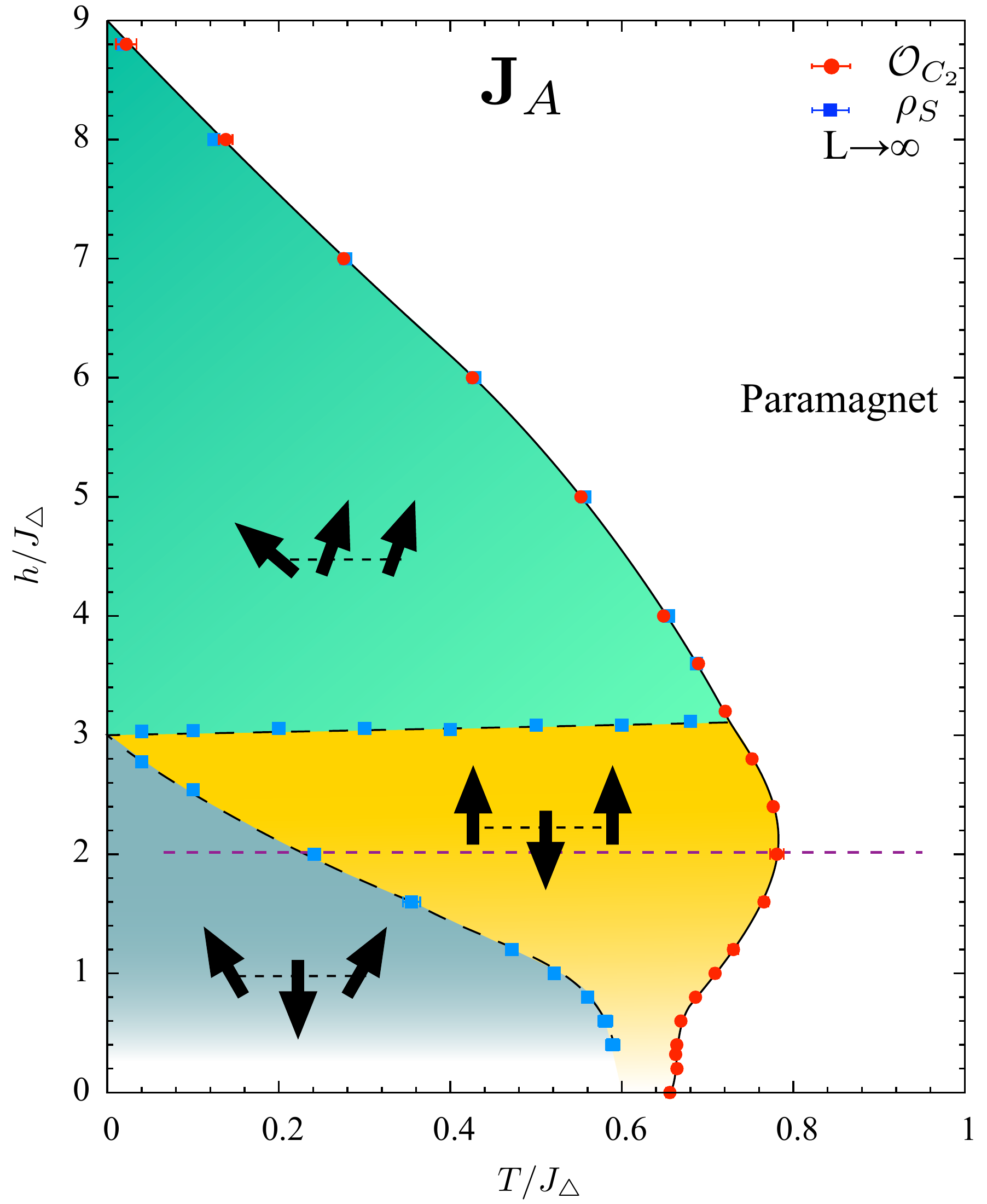}
\caption{
(Color online).
Phase diagram of a square--lattice frustrated ferromagnet in applied magnetic field,
exhibiting the same phases as found in the Heisenberg antiferromagnet
on a triangular lattice [\onlinecite{Seabra2011a}] ---   a coplanar ``Y--state'', interpolating
to ``120--degree'' order at vanishing magnetisation; a collinear $1/3$--magnetisation
plateau [cf.~Fig.~\ref{fig:novel.phases}(a)]; and a coplanar 2:1 canted state.
Phase boundaries are taken from classical Monte Carlo simulations of
$\mathcal{H}^{\sf FFM}_\square$~[Eq.~(\ref{eq:Hex})], for the parameter set
${\bb J}_A$~[Eq.~(\ref{eq:def.JA})], and scaled to the thermodynamic limit,
as described in Section~\ref{section:simulations.for.JA}.
Temperature and magnetic field
measured in units of $J_\triangle$~[Eq.~(\ref{eq:def.J.triangle})].}
\label{fig:JTphase}
\end{figure}

In this Article we explore some of the novel phases which arise in
a simple example of a frustrated ferromagnet
--- a Heisenberg model on a square lattice, in which ferromagnetic
1$^{st}$--neighbour interactions compete with antiferromagnetic
2$^{nd}$-- and 3$^{rd}$-- neighbour exchange.
Using a combination of classical Monte Carlo simulation and
spin--wave theory, we establish the phase diagram of this
model as a function of temperature and magnetic field,
for two different sets of exchange parameters.
In the process, we uncover a number of phases
not usually associated with square--lattice magnets.
These include a collinear $1/3$--magnetisation plateau, a spin liquid
with finite magnetisation, and a crystal composed of magnetic vortices.
Illustrations of these phases, and of the phase diagram for one
of the parameter sets, are shown in Fig.~\ref{fig:novel.phases}
and Fig.~\ref{fig:JTphase}, respectively.


\begin{figure}[t]
\includegraphics[width=\columnwidth/2]{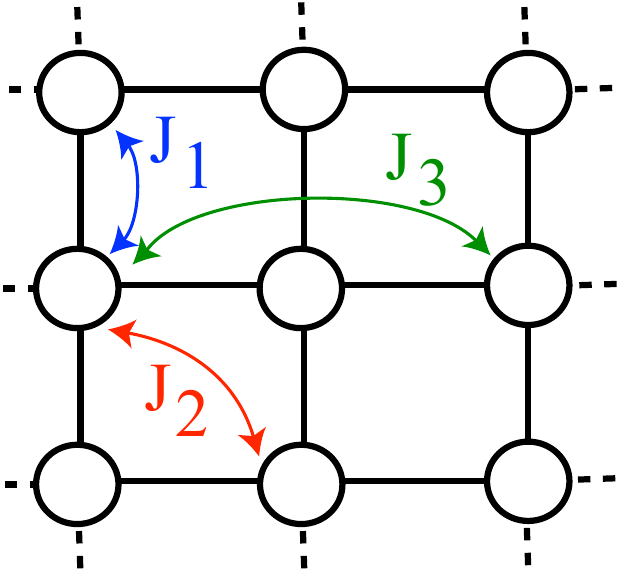}
\caption{
(Color online).
Exchange interactions up to \mbox{3$^{rd}$--neighbour} on the square lattice.
In this Article we consider
a Heisenberg model $\mathcal{H}^{\sf FFM}_\square$~[Eq.~(\ref{eq:Hex})]
where ferromagnetic \mbox{1$^{st}$--neighbour} exchange $J_1$ competes with antiferromagnetic
2$^{nd}$-- and \mbox{3$^{rd}$--neighbour} exchanges $J_2$ and $J_3$.
\label{fig:exchange.interactions}
}
\end{figure}


The model we consider is a natural generalisation of the ``\mbox{$J_1$--$J_2$}''
model, which describes competing Heisenberg exchange interactions on the
1$^{st}$-- and  2$^{nd}$--neighbour bonds of a square lattice.
The \mbox{$J_1$--$J_2$} model has a long and interesting history, and remains
one of the paradigmatic examples of a frustrated magnet~\cite{Misguich2004book}.
Much of the original interest in this model was driven by the
possibility that competing antiferromagnetic exchange could stabilise
a quantum spin--liquid, a question which continues to inspire new
research~\cite{Jiang2012,Hu2013,Doretto2014,Gong2014}.
However the \mbox{$J_1$--$J_2$} model is also significant as a working
model of magnetism in iron pnictides~\cite{Si2008,smerald10,dai15},
and as one of the simplest possible examples of a frustrated ferromagnet.
In particular, where ferromagnetic
1$^{st}$--neighbour exchange competes with antiferromagnetic
2$^{nd}$--neighbour exchange, the spin--$1/2$ \mbox{$J_1$--$J_2$} model can
support spin--nematic order~\cite{Shannon2004,Shannon2006,Schmidt2007,Thalmeier2008,Zhitomirsky2010,
Richter2010a,Shindou2011,Smerald2015,Tsirlin2009}.


Extending frustrated spin models to include further--neighbour interactions typically
results in classical ground states with incommensurate, helical order.
For large values of spin, the leading effect of quantum and/or thermal
fluctuations is a small correction to the pitch of the helix, leading to
small, quantitative, changes in phase boundaries~\cite{Diep1989,Rastelli1985,Ueda2009}.
In the case of the \mbox{$J_1$--$J_2$}--$J_3$ Heisenberg model on
a square lattice, these expectations are borne out by linear spin-wave
theory~\cite{Rastelli1979,Rastelli1986,Chubukov1984}.
However exact diagonalisation calculations for \mbox{$S=1/2$} suggest a richer
phase diagram, including a number of novel ground states~\cite{Sindzingre2009,Sindzingre2010}.
And even at a classical level, other models, such as the \mbox{$J_1$--$J_2$}--$J_3$
Heisenberg model on a honeycomb lattice, can support a much richer behaviour\cite{Okumura2010,Rosales2013}.


One of the possible alternatives to simple helical order are \mbox{multiple-Q} states,
composed of a coherent superposition of states with different ordering vectors.
In general, these states are not compatible with the fixed spin length constraint $|\mathbf{S}|=1$,
and therefore do not belong to ground--state manifolds.
However, there are a number of recent interesting cases where these phases are
stabilised by the interplay between frustration and thermal fluctuations, such as the classical
pyrochlore antiferromagnet~\cite{Okubo2011} or the classical extended
triangular-lattice antiferromagnet~\cite{Okubo2012}.
Interestingly, multiple-Q states also arise in the description of spin
textures composed of crystals of topological defects.
These are the celebrated case of skyrmions lattices in chiral magnets such
as MnSi~\cite{Muhlbauer2009,Schulz2012}, magnetic-vortex lattices in a generic class of Mott insulators~\cite{Kamiya2014}, or skyrmion crystals in the antiferromagnetic triangular lattice~\cite{Rosales2015}.
Magnetic-vortex lattices in insulating systems seem  to be especially uncommon outside the realm of superconducting
systems~\cite{deGennes-book,Blatter1994}.


The picture which emerges from the square--lattice frustrated ferromagnet
studied in this Article is interesting for a number of reasons.
Just as in frustrated antiferromagnets, the interplay between ground state degeneracy
and fluctuations leads to a rich behaviour in applied magnetic field --- cf. Fig.~\ref{fig:JTphase}.
However, unlike those antiferromagnets where the frustration comes from the geometry
of the lattice, the model under study here can be tuned to exhibit a wide range of different magnetic
phenomena, many of them unexpected on a square lattice.
This is particularly true when parameters are chosen so to place the model at the
border  between competing forms of order.

The remainder of this Article is structured as follows:

In Sec.~\ref{section:model.and.methods} we introduce the model and the methods
used to study it.
On the basis of known results for the classical ground--state
we single out two parameter sets; one
which corresponds, at a mean--field level, to a triangular--lattice antiferromagnet;
and another
for which the classical ground--state is highly degenerate.


In Sec.~\ref{section:simulations.for.JA} we present Monte Carlo simulation results
for the first of these parameter sets.
We demonstrate that the phase diagram as a function of magnetic field and temperature
is almost identical to that of the Heisenberg antiferromagnet on a triangular lattice,
and includes its celebrated \mbox{$1/3$--magnetisation} plateau, now translated to
three--sublattice order on a square lattice.
These results are summarised in Fig.~\ref{fig:JTphase}.


In Sec.~\ref{section:simulations.for.JB} we present Monte Carlo simulation results
for the second of these parameter sets.
We establish the phase diagram of a finite--size cluster as a function of magnetic
field and temperature, and present evidence for both a finite--temperature classical
spin--liquid and, at low temperature and low values of magnetic field, a multiple--Q state
with the character of a vortex crystal.
These results are summarised in Fig.~\ref{fig:JRphase}.


In Sec.~\ref{section:discussion} we discuss how these results relate to published
work on frustrated antiferromagnets, how quantum effects might enter into the problem,
and where these novel phases might be realised in experiment.


Finally, in Sec.~\ref{section:conclusions}, we conclude with a brief
summary of the results and remaining open questions.


Technical details of associated spin--wave calculations, used to confirm the results of
Monte Carlo simulations, are discussed in a small number of Appendices.


Appendix~\ref{section:Appendix_spin_wave} develops the mathematical formalism needed
to carry out both a classical low--temperature expansion, and a linear spin--wave expansion,
about the classical ground states found in this model. 	


Appendix~\ref{section:selection_1Q_zero_h} applies this analysis to the ground state 
manifold for the second parameter set, and shows that thermal fluctuations select a spiral 
state at low temperatures and zero field, as suggested from Monte Carlo simulations.


Appendix~\ref{section:selection_1Q_finite_h} analyses how, 
for the second parameter set, a canted $Y$ state eventually prevails over a conical 
version of the spiral state when a magnetic field is applied.
The vortex crystal is not favoured at very low temperatures, again in agreement
with the Monte Carlo results.


Appendix~\ref{section:selection_quantum_model} discusses,
from linear spin--wave calculations of the quantum corrections
to the ground-state energy, 
how the 
classical phase diagram for the second parameter set changes
in the quantum model at low magnetic fields and temperatures.

\section{Model, Method and Order Parameters}
\label{section:model.and.methods}

\subsection{The \mbox{$J_1$--$J_2$}--$J_3$ Heisenberg model on a square lattice}
\label{section:model}

In this Article, we consider one of the simplest possible prototypes for a
frustrated ferromagnet, the Heisenberg model on a square lattice, with competing
2$^{nd}$-- and 3$^{rd}$--neighbour exchange
\begin{eqnarray}
\mathcal{H}^{\sf FFM}_\square
   &=&  J_1 \sum_{\langle ij \rangle_1} \mathbf{S}_i \cdot \mathbf{S}_j
            + J_2 \sum_{\langle ij \rangle_2} \mathbf{S}_i \cdot \mathbf{S}_j \nonumber\\
   &&    \quad + J_3 \sum_{\langle ij \rangle_3} \mathbf{S}_i \cdot \mathbf{S}_j  - h \sum_i S_i^z \; .
\label{eq:Hex}
\end{eqnarray}
Here $\mathbf{S}_i$ is a classical spin with $|\mathbf{S}_i|^2=1$, 
the sum $\sum_{\langle ij \rangle_n}$ runs over
the n$^{th}$--neighbour bonds of a square lattice, as illustrated in
Fig.~\ref{fig:exchange.interactions}, and $h$ is an applied magnetic field.
We restrict ourselves to the case where 1$^{st}$--neighbour exchange
$J_1$ is ferromagnetic, while further--neighbour exchange $J_2$ and $J_3$
are antiferromagnetic, i.e.
\begin{eqnarray}
J_1 < 0 \quad , \quad J_2 >  0 \quad , \quad  J_3 > 0 \; .
\label{eq:conditions.on.J}
\end{eqnarray}
The exchange integrals $J_1$, $J_2$, $J_3$ define a 3--dimensional parameter space,
and for many purposes, it is  convenient to represent them as a vector
\begin{eqnarray}
{\bf J} = (J_1,\ J_2,\ J_3) \; .
\label{eq:J.as.vector}
\end{eqnarray}


The classical ground--state phase diagram of $\mathcal{H}^{\sf FFM}_\square$~[Eq.~(\ref{eq:Hex})],
together with its spin--wave excitations, was studied in a series of papers
by Rastelli~et~al.~\cite{Rastelli1979,Rastelli1986} and Chubukov~\cite{Chubukov1984}.
In the absence of magnetic field, allowing for the leading effect of fluctuations,
all ground states are found to be coplanar spirals characterised
by a wave vector
\begin{eqnarray}
{\bf Q} = (Q_x, Q_y) \; .
\label{eq:def.Q}
\end{eqnarray}
This wave vector can be determined by
minimising the Fourier transform of the interactions
\begin{eqnarray}
J({\bf q})
   &=& 2J_1(\cos q_x +\cos q_y)
   + 4J_2 \cos q_x \cos q_y \nonumber \\
   && \qquad +  2J_3 (\cos2 q_x+\cos 2 q_y) \; .
\label{eq:FT.interactions}
\end{eqnarray}
For ferromagnetic $J_1$ [cf. Eq.~(\ref{eq:conditions.on.J})], there
are four distinct cases~:
\begin{enumerate}

\item  a uniform ferromagnetic (FM) phase with
\begin{eqnarray}
{\bb Q}^{\sf FM} = (0,0).
\label{eq:def.Q.FM}
\end{eqnarray}

\item a two--sublattice collinear antiferromagnetic (CAF) phase with
\begin{eqnarray}
{\bb Q}^{\sf CAF} = (\pi,0) \; \text{or} \; (0,\pi).
\label{eq:def.Q.CAF}
\end{eqnarray}

\item a family of one--dimensional (1D) spirals with
\begin{eqnarray}
{\bb Q}^{\sf 1D}
    &=& (Q_{\sf 1D}, 0)  \; \text{or} \;  (0, Q_{\sf 1D}) \; , \nonumber\\
&& \qquad \cos Q_{\sf 1D} = -\frac{J_1+ 2J_2}{4J_3} \; .
\label{eq:def.Q.1D}
\end{eqnarray}

\item  a family of two--dimensional (2D) spirals with
\begin{eqnarray}
{\bb Q}^{\sf 2D}
    &=& (Q_{\sf 2D}, Q_{\sf 2D}) \; , \nonumber\\
&& \qquad  \cos Q_{\sf 2D}=  -\frac{J_1}{2J_2+4J_3} .
\label{eq:def.Q.2D}
\end{eqnarray}

\end{enumerate}
These ground states, and the range of parameters for which they occur,
are illustrated in Fig.~\ref{fig:classical.ground.state.phase.diagram}.

\subsection{Three--sublattice order on the square lattice, and its connection
                   with a triangular--lattice antiferromagnet}
\label{sec:mean.field.triangles}

\begin{figure}[t]
\includegraphics[width=8cm]{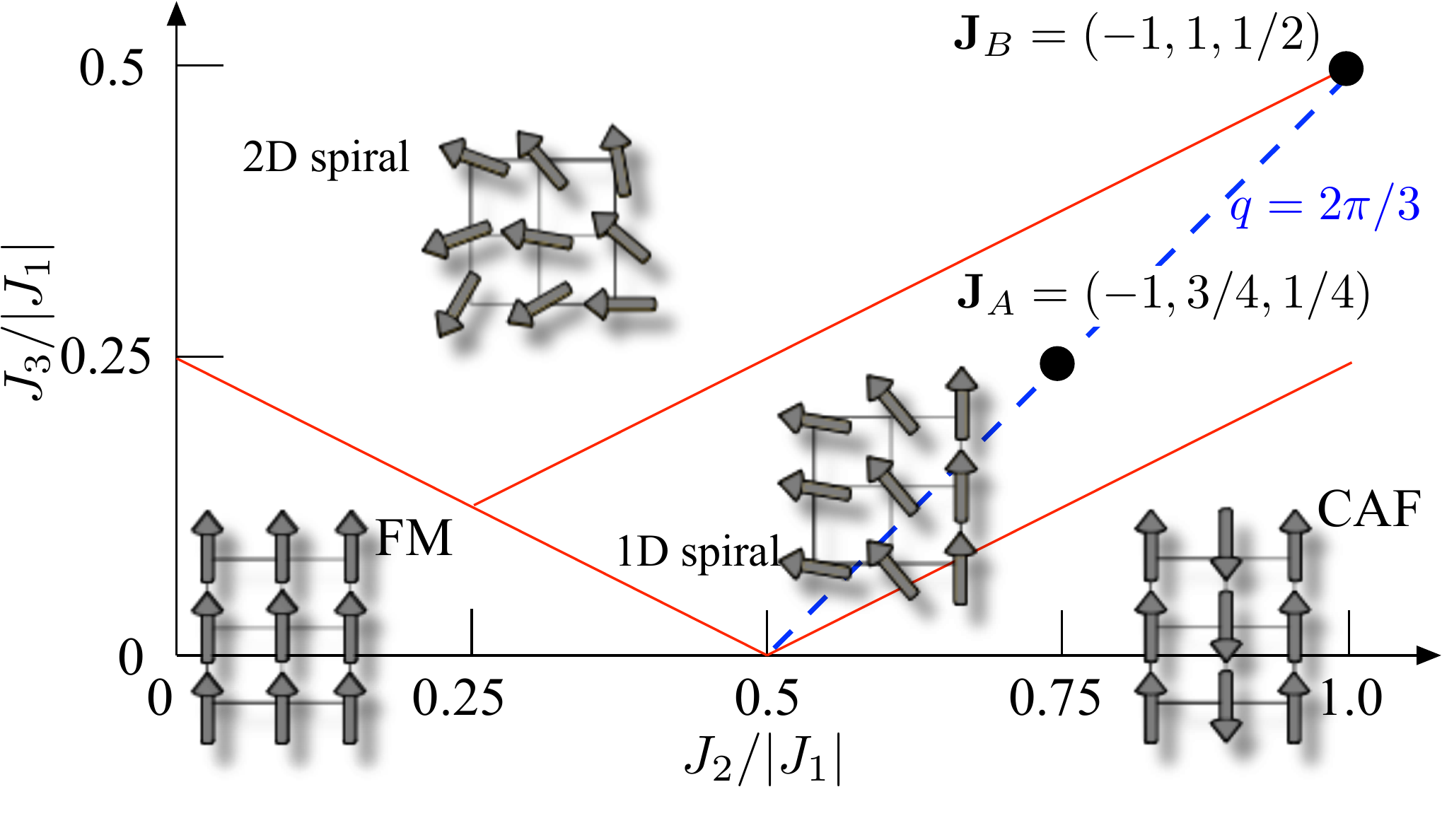}
\caption{ %
(Color online).
Classical ground--state phase diagram of the frustrated Heisenberg ferromagnet on
on a square lattice, $\mathcal{H}^{\sf FFM}_\square$~[Eq.~(\ref{eq:Hex})], as a function of
competing antiferromagnetic exchange $J_2$ and $J_3$, following~\linecite{Rastelli1979}.
Ground states in the absence of magnetic field comprise a collinear ferromagnet
(FM), a collinear antiferromagnet (CAF), a one--dimensional coplanar spiral (1D spiral)
and a two--dimensional coplanar spiral (2D spiral), as defined
in Eqs.~(\ref{eq:def.Q.FM}--\ref{eq:def.Q.2D}).
The two parameter sets studied in this Article, ${\bb J}_A$ [Eq.~(\ref{eq:def.JA})]
and ${\bb J}_B$  [Eq.~(\ref{eq:def.JB})],
are labeled with black dots.
A blue dashed line indicates parameters for which the ground state is a 1D spiral
with three-sublattice order [Eq.~(\ref{eq:def.triangle.line})].
\label{fig:classical.ground.state.phase.diagram}
}
\end{figure}

In general, the one--dimensional spiral ground states of
$\mathcal{H}^{\sf FFM}_\square$~[Eq.~(\ref{eq:Hex})] are incommensurate, with a wave vector
${\bb Q}^{\sf 1D}$ [Eq.~(\ref{eq:def.Q.1D})]
which interpolates smoothly from one phase
to another --- e.g. from ${\bb Q}^{\sf FM}$~[Eq.~(\ref{eq:def.Q.FM})]
to ${\bb Q}^{\sf CAF}$~[Eq.~(\ref{eq:def.Q.CAF})].
For certain choices of exchange parameters, however, ${\bb Q}^{\sf 1D}$ takes
on commensurate values.
One particular case, occurring for %
\begin{eqnarray}
J_1 +  2 J_2 - 2 J_3 = 0
   \quad , \quad
        [\ 0 < 2 J_3 <   |J_1| \ ] \; ,
\label{eq:def.triangle.line}
\end{eqnarray}
[cf. the blue dashed line in Fig.~\ref{fig:classical.ground.state.phase.diagram}], is
\begin{eqnarray}
{\bb Q}^{\sf 1D}_{\sf 3sub}
    &=& \left( \frac{2\pi}{3}, 0 \right)
    \; \text{or} \;
    \left( 0, \frac{2\pi}{3} \right)  \; .
\label{eq:def.Q.1D.triangle}
\end{eqnarray}
In this case the ground state of the frustrated ferromagnet
$\mathcal{H}^{\sf FFM}_\square$~[Eq.~(\ref{eq:Hex})]
is composed of stripes of spins, 
with three--sublattice order.


Three--sublattice order also occurs in one of the %
paradigmatic examples of frustrated antiferromagnetism, the Heisenberg antiferromagnet
on the {\it triangular} lattice
\begin{eqnarray}
\mathcal{H}^{\sf AF}_\triangle
   &=&  J \sum_{\langle ij \rangle} \mathbf{S}_i \cdot \mathbf{S}_j  - h \sum_i S^z_i
   \quad , \quad J > 0 \; .
\label{eq:H.trianglar.AF}
\end{eqnarray}
First studied as a potential route to a quantum spin liquid~\cite{Anderson1973},
this model has a long and distinguished history as a testing ground for ideas
about classical and quantum magnets~\cite{Starykh2014,Chen2013}.
Its behaviour in magnetic field, in particular, where a collinear  \mbox{$1/3$--magnetisation} plateau
is selected by fluctuations from a degenerate set of classical ground states~\cite{Kawamura1985,Seabra2011a,Gvozdikova2011},
has come to be seen as one of the paradigms for frustrated magnets.
%


At a mean--field level,
for a 1D spiral state with wave vector ${\bb Q}^{\sf 1D}_{\sf 3sub}$~[Eq.~(\ref{eq:def.Q.1D.triangle})],
$\mathcal{H}^{\sf FFM}_\square$~[Eq.~(\ref{eq:Hex})] is equivalent to
$\mathcal{H}^{\sf AF}_\triangle$~[Eq.~(\ref{eq:H.trianglar.AF})], with an effective interaction
\begin{eqnarray}
J = J_\triangle &=& \frac{J_1+2J_2+J_3}{3}  \; .
\label{eq:def.J.triangle}
\end{eqnarray}
As a consequence, the frustrated ferromagnet, $\mathcal{H}^{\sf FFM}_\square$, inherits
many of the special properties of the Heisenberg antiferromagnet on a triangular lattice.
In particular, the classical ground states of
$\mathcal{H}^{\sf AF}_\triangle$
can be found by rewriting the model as
\begin{eqnarray}
\mathcal{H}^{\sf AF}_\triangle
   &=&   \frac{9 }{2} J_\triangle \sum_\triangle \left[
   \left({\bb m}_\triangle
   - \frac{h }{9 J_\triangle} \hat{\bb e}_z \right)^2
    - \frac{S^2}{3}
    - \frac{ h^2}{81 J^2_\triangle}
    \right] \; , \nonumber\\
    \label{eq:H.triangle.triangle}
\end{eqnarray}
where $\hat{\bb e}_z$ is a unit vector in the direction of the magnetic field,
the sum $\sum_\triangle$ runs over all triangles
in the lattice, and
\begin{eqnarray}
{\bb m}_\triangle
   &=&   \frac{1}{3} \sum_{i \in \triangle} {\bb S}_i
\end{eqnarray}
is the magnetisation {\it per spin} in a given triangle.


It follows from Eq.~(\ref{eq:H.triangle.triangle}) that {\it any} state for which
\begin{eqnarray}
{\bb m}_\triangle \equiv \frac{h}{9 J_\triangle}   \hat{\bb e}_z
   \quad \forall \quad \triangle \in \text{lattice} \; ,
\label{eq:m.of.h}
\end{eqnarray}
is a classical ground state of $\mathcal{H}^{\sf AF}_\triangle$,
and that the classical ground state interpolates
to saturation (full magnetisation) for a magnetic field
\begin{eqnarray}
h_{\sf sat}  &=&  9 J_\triangle \; .
\label{eq:def.h.sat}
\end{eqnarray}
At a mean--field level, both of these properties carry over to three-sublattice
ground states of $\mathcal{H}^{\sf FFM}_\square$~[Eq.~(\ref{eq:Hex})].


However this is by no means the end of the story --- except at saturation, the
mean--field constraint on ${\bb m}_\triangle$~[Eq.~(\ref{eq:m.of.h})], {\it does not}
uniquely constrain the ground state, and fluctuations play a crucial role in
establishing order.
In the case of the Heisenberg antiferromagnet on a triangular lattice,
$\mathcal{H}^{\sf AF}_\triangle$ [Eq.~(\ref{eq:H.trianglar.AF})],
it is known that thermal~\cite{Kawamura1985,Henley1989,Seabra2011a}
and/or quantum\cite{Chubukov_Golosov_1991}
fluctuations act on the large family of degenerate
three--sublattice ground states %
to select~:
\begin{itemize}

\item  a coplanar  \mbox{``120--degree''} state with zero magnetisation,
          for $h=0$.

\item  a coplanar ``Y--state'' with finite magnetisation, interpolating to
          120--degree order for $h \to 0$.

\item a collinear ``uud'' state associated with
         a \mbox{$1/3$--magnetisation} plateau for intermediate values of $h$.

\item a coplanar 2:1 canted state, interpolating between the \mbox{$1/3$--magnetisation}
         plateau and the saturated state for $h \to h_{\sf sat}$.

\end{itemize}
The associated classical phase diagram is shown in Fig.~1 of Ref.~\onlinecite{Seabra2011a}.
Since strict long--range order of phases which break a continuous symmetry
is forbidden at finite temperature in two dimensions by the Mermin--Wagner theorem,
the coplanar phases should be understood as algebraically correlated.


In Sec.~\ref{section:simulations.for.JA} we use classical Monte Carlo simulation to explore how the
equivalent three--sublattice order in the square--lattice frustrated ferromagnet
$\mathcal{H}^{\sf FFM}_\square$~[Eq.~(\ref{eq:Hex})] evolves as a function of temperature and magnetic field.
For these purposes, we select the parameter set
\begin{eqnarray}
{\bb J}_A = (J_1,\ J_2,\ J_3) = (-1,\ 3/4,\ 1/4),
\label{eq:def.JA}
\end{eqnarray}
marked with a black dot in Fig.~\ref{fig:classical.ground.state.phase.diagram},
as representative of parameters with a three--sublattice ground state
[cf.~Eq.~(\ref{eq:def.triangle.line})].
This in turn sets a characteristic scale for temperature and magnetic field,
through Eq.~(\ref{eq:def.J.triangle}).

\subsection{Highly--degenerate manifold of classical ground states}
\label{subsec:ring-equations}

An enlarged ground state manifold occurs
where the 1D and 2D spirals meet, for %
\begin{eqnarray}
J_2  - 2 J_3 = 0
   \quad \forall \quad
          |J_1|  <  4 J_2 <  4 |J_1|  \; ,
\label{eq:def.degenerate.line}
\end{eqnarray}
[cf. Fig.~\ref{fig:classical.ground.state.phase.diagram}].
On this phase boundary, the classical ground states of  $\mathcal{H}^{\sf FFM}_\square$
are 2D spirals with wave vector
\begin{eqnarray}
{\bb Q}^{\sf ring} &=& ( Q^{\sf ring}_x,\  Q^{\sf ring}_y)  \; , \nonumber\\
&& Q^{\sf ring}_y = \pm \arccos\Big(\frac{-J_1 }{2J_2}-\cos Q^{\sf ring}_x\Big) \; .
\label{eq:ring-qx-qy}
\end{eqnarray}
which interpolates from the 1D to the 2D spiral.
This set of ${\bb Q}$ forms a ring in reciprocal space, centred on
\begin{eqnarray}
{\mathbf \Gamma} = (0,\ 0) \; ,
\end{eqnarray}
and defines a highly degenerate manifold of states.


In Sec.~\ref{section:simulations.for.JB} we use classical Monte Carlo simulation to explore
the consequences of this enlarged ground--state degeneracy at finite temperature and magnetic
field.
We consider the limiting case where the line Eq.~(\ref{eq:def.triangle.line}), corresponding to a 1D spiral
with three-sublattice order, terminates at the phase boundary between 1D and 2D spirals at
\begin{eqnarray}
{\bb J}_B   = (J_1,\ J_2,\ J_3) = (-1,\ 1,\ 1/2) \; .
\label{eq:def.JB}
\end{eqnarray}
This parameter set is denoted as a black dot in
Fig.~\ref{fig:classical.ground.state.phase.diagram}.
The set of ground--state wave vectors
${\bb Q}^{\sf ring}$ [Eq.~(\ref{eq:ring-qx-qy})] at ${\bb J}_B$ is
shown as a dashed line in Fig.~\ref{fig:novel.phases}(b).

\subsection{Monte Carlo simulation method}

In Sec.~\ref{section:simulations.for.JA} and Sec.~\ref{section:simulations.for.JB}
of this paper, we use large--scale Monte Carlo simulation to study the finite--temperature 
properties of the square lattice frustrated ferromagnet 
$\mathcal{H}^{\sf FFM}_\square$~[Eq.~(\ref{eq:Hex})] in applied magnetic field.
Simulations were performed using parallel tempering~\cite{Hukushima1996},
using 48 to 80 replicas (temperatures).
Simulations were carried out for square clusters of
\begin{eqnarray}
N = L \times L
\end{eqnarray}
spins, with periodic boundary conditions.
The linear sizes $L$ were chosen to be  commensurate with possible ${\bb Q}$
wave vectors, in the range $60 \leq L \leq 180$.
Typical simulations involved 2$\times$10$^6$ steps, half of which were discarded for thermalisation.
At every 10 steps there was an attempt at exchanging replicas at neighbouring temperatures.
Energy scales (field and temperature) were normalised to $J_\triangle$~[Eq.~(\ref{eq:def.J.triangle})],
for easy comparison with the triangular lattice antiferromagnet~\cite{Seabra2011a}.


The simulations for the parameter set${\bb J}_B$~[Eq.~(\ref{eq:def.JB})],
presented in Sec.~\ref{section:simulations.for.JB}, become very challenging at low
temperatures, especially for low values of magnetic field, where several different
phases compete.
In this case we performed different simulations runs, with and without parallel tempering,
where the initial state was either one of the candidate phases at $T=0$, or  a
configuration composed of domains of different phases, in order to ascertain their
relative stability.

\subsection{A short catalogue of order parameters and correlation functions}

Experience of simulating the Heisenberg antiferromagnet on a triangular lattice
[Ref.~\onlinecite{Seabra2011a}], together with the symmetry of the ordered phases found in the absence
of magnetic field [cf. Fig.~\ref{fig:classical.ground.state.phase.diagram}],
suggests a number of order parameters and correlation functions likely to be of use in determining
the behaviour of $\mathcal{H}^{\sf FFM}_\square$~[Eq.~(\ref{eq:Hex})] in applied magnetic field.


The 1D spiral phase in Fig.~\ref{fig:classical.ground.state.phase.diagram} breaks the four--fold
rotational symmetry of the square lattice  from $C_4$ down to $C_2$, by choosing to orient the
stripes in vertical or horizontal direction.
We analyse this with the following  order parameter
\begin{eqnarray}
%
%
\mathcal{O}_{C_2} &=& \Big\langle  | \phi_{\bf x} |   \Big\rangle
\label{eq:o-c2-3} \; , \\
\phi_{\sf {\bf x}} &=& \frac{9}{32N} \sum_i \mathbf{S}_i \cdot (\mathbf{S}_{i+\xvec}-\mathbf{S}_{i+\yvec})
\label{eq:o-c2-1}. \;
\end{eqnarray}
The sum over $i$ runs over all $N = L^2$ lattice sites and
the lattice unit vectors are denoted as  $\hat{\bf x}=(1,0)$ and  $\yvec=(0,1)$.


In the presence of field, the three-sublattice ordered states break the translational 
symmetry of the lattice along the $S^z$ direction.
This can be measured by an order parameter based on a two-dimensional irreducible
representation of $C_3 \cong \mathds{Z}_3$, cf. the triangular-lattice case~\cite{Seabra2011a}
\begin{eqnarray}
\psi_{\sf {\bf x}, 1}^z && =\frac{3}{\sqrt{6}N}  \sum_i 2S^z_i +2 S^z_{i+\xvec}-4S_{i+2\xvec}^z		 \label{eq:o-c3-1},  \\
\psi_{\sf {\bf x}, 2}^z &=&-\frac{3}{\sqrt{2}N} \sum_i 2 S^z_{i+\xvec}-2S^z_i. 					 \label{eq:o-c3-2}
\end{eqnarray}
where the sum over $i$ runs over the $N/3$ ``3-spin stripes'', each equivalent to an elementary triangle.
In order to account for the two ways of  breaking the $C_2$ symmetry, the $\mathds{Z}_3$ order
parameter must be measured along both directions of the lattice, i.e. $\xvec\rightarrow\yvec$
in Eqs.~(\ref{eq:o-c3-1}) and (\ref{eq:o-c3-2}).
The resulting final order parameter is
\begin{eqnarray}
\mathcal{O}^z_{\mathds{Z}_3}
   &=& \Bigg\langle
   \Big[ |\psi^z_{\sf {\bf x}, 1}|^2
    + |\psi^z_{\sf {\bf x}, 2}|^2
    + |\psi^z_{\sf {\bf y}, 1}|^2
    + |\psi^z_{\sf {\bf y}, 2}|^2	
    \Big]^{1/2}  \Bigg\rangle \; .
    \nonumber\\
\label{eq:o-c3-3}
\end{eqnarray}
The susceptibility associated with each order parameter $\mathcal{O}$ is defined as
\begin{eqnarray}
\chi = N\frac{\langle |\mathcal{O}|^2 \rangle - \langle |\mathcal{O}| \rangle^2}{T}.
\end{eqnarray}


In the presence of magnetic field, quasi--long--range order can develop in the transverse
components of spin
\begin{eqnarray}
\mathbf{S}_i^\perp = ( S_i^x, S_i^y ) \; .
\label{eq:def.S.perp}
\end{eqnarray}
This is captured by the spin stiffness $\rho_S$,
see e.g. Refs.~\onlinecite{Seabra2011,Seabra2011a} and references therein,
\begin{eqnarray}
\rho_s[\textbf{\^{e}}]=&&  \frac{1  }{N}  \Bigg\langle \sum_\delta \
J_\delta \sum_{{\langle i,j \rangle}_\delta}
(\textbf{\^{e}} \cdot \mathbf{r}_{ij})^2  \mathbf{S}^\perp_i \cdot \mathbf{S}^\perp_j
\Bigg\rangle
 \nonumber \\
&&
- \frac{1}{NT} \Bigg\langle   \Big(  \sum_\delta	J_\delta \sum_{{\langle i,j \rangle}_\delta}
(\textbf{\^{e}} \cdot \mathbf{r}_{ij})  \mathbf{S}^\perp_i
 \times \mathbf{S}^\perp_j
 \Big)^2     \Bigg\rangle,
\label{eq:rhos-3}
\end{eqnarray}
which is  averaged over the two lattice directions \mbox{$\mathbf{\hat{e}}=\{ \xvec , \yvec \}$}.


The specific heat, defined as
\begin{eqnarray}
C_h = \frac{1}{N}\frac{\langle E \rangle - \langle E \rangle^2}{T^2},
\end{eqnarray}
where $E$ is the total internal energy, is also helpful in tracking phase transitions.


Last, but not least, the momentum-resolved spin structure factor
\begin{eqnarray}
\mathcal{S}(\mathbf{q}) = \Big\langle \frac{1}{N}
\Big| \sum_i\mathbf{S}_i\exp(-i\mathbf{q}\cdot\mathbf{r}_i)\Big|^2 \Big\rangle.
\label{eq:S(q)-def}
\end{eqnarray}
is of considerable importance for characterising the spin--liquid phase studied
in Sec.~\ref{section:simulations.for.JB}.
Where appropriate, we analyse separately the structure factors
for the longitudinal and transverse components of spin
\begin{eqnarray}
\mathcal{S}^z (\mathbf{q}) &=& \Big\langle \frac{1}{N} \Big|
\sum_i S^z_i
\exp(-i\mathbf{q}\cdot\mathbf{r}_i)\Big|^2 \Big\rangle \; ,
\label{eq:S.longitundinal.of.q}
\\
\mathcal{S}^\perp (\mathbf{q}) &=& \Big\langle \frac{1}{N} \Big|
\sum_i \mathbf{S}^\perp_i\exp(-i\mathbf{q}\cdot\mathbf{r}_i)\Big|^2 \Big\rangle \; .
\label{eq:S.perpendicular.of.q}
\end{eqnarray}
where ${\bb S}^\perp$ is defined in Eq.~(\ref{eq:def.S.perp}).


\begin{figure}[t!]
\centering
\includegraphics[width=\columnwidth]{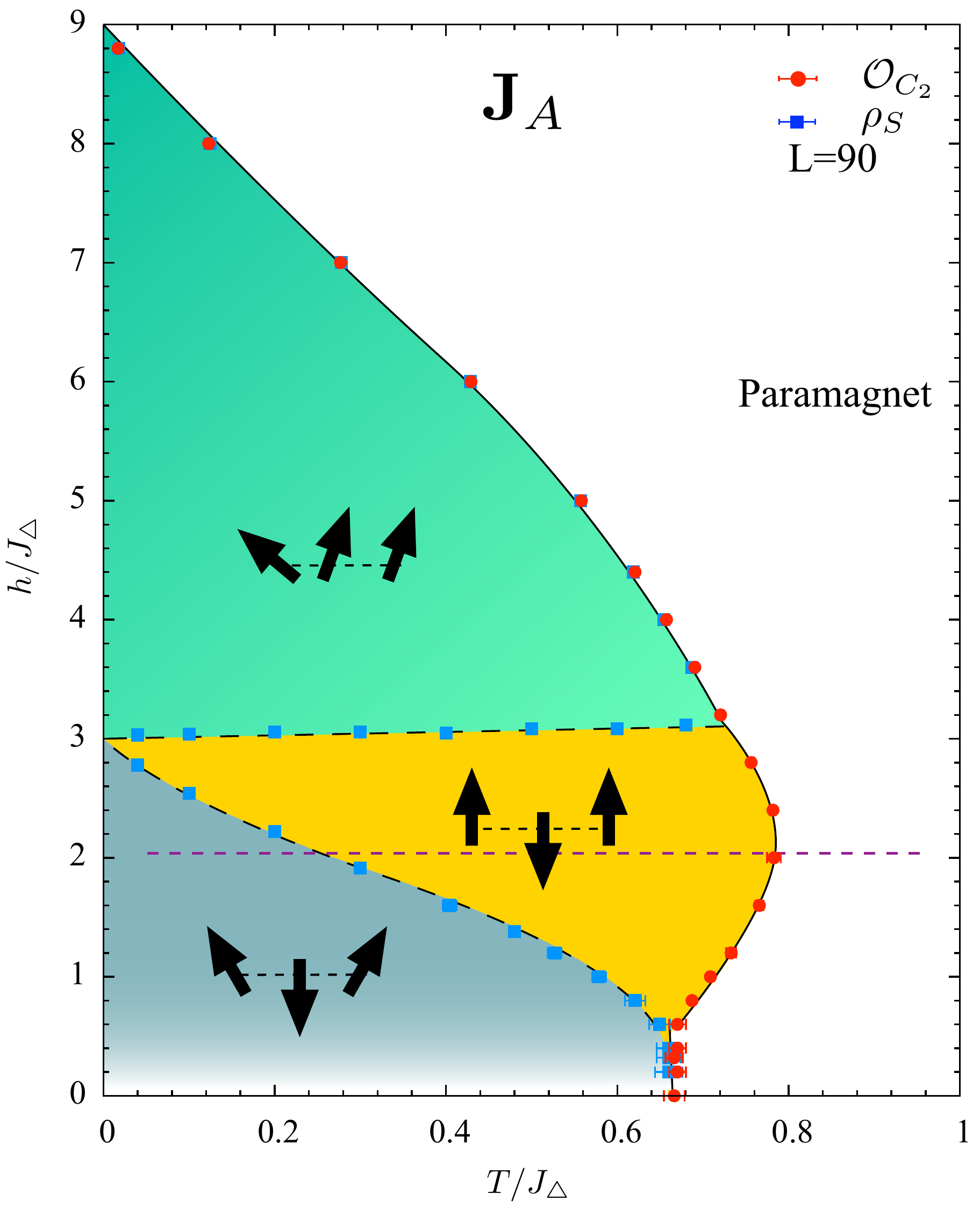}
\caption{%
(Color online).
Finite--size phase diagram of a square--lattice frustrated ferromagnet in applied magnetic field.
The phases found ---  a coplanar ``Y--state'';  a collinear $1/3$--magnetisation plateau;
and a coplanar 2:1 canted state --- and the structure of the phase diagram, closely parallel
finite--size results for the Heisenberg antiferromagnet on a triangular lattice~\cite{Seabra2011a}.
In both cases, finite--size effects strongly renormalise the temperature associated
with the transition from the collinear $1/3$--magnetisation plateau into the Y--state.
Results are taken from classical Monte Carlo simulations of
$\mathcal{H}^{\sf FFM}_\square$~[Eq.~(\ref{eq:Hex})] for the parameter set
${\bb J}_A$~[Eq.~(\ref{eq:def.JA})].
Phase boundaries were extracted from anomalies in
the order parameter
associated with lattice rotations~[Eq.~(\ref{eq:o-c2-3})]
and spin stiffness $\rho_S$~[Eq.~(\ref{eq:def.S.perp})],
for a cluster of $N = 90^2= 8100$ spins.
Temperature and magnetic field are measured in units of
$J_\triangle$~[Eq.~(\ref{eq:def.J.triangle})].
The horizontal dashed line
corresponds to the temperature cut used in
Fig.~\ref{fig:JT-field}.
}
\label{fig:JTphaseL90}
\end{figure}


\begin{figure}[t]
\centering
\includegraphics[width=8cm]{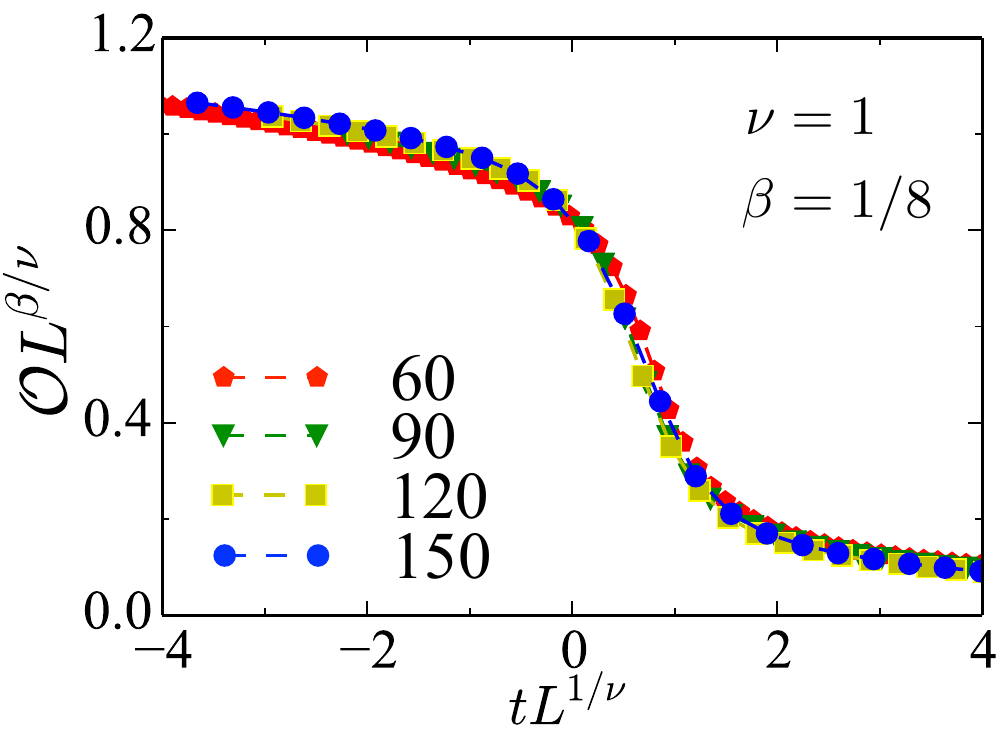}
\caption{
(Color online).
Finite--temperature phase transition into a three-sublattice, ``120--degree'' ground state,
breaking lattice rotation symmetry.
The order parameter $\mathcal{O}_{C_2}$~[Eq.~(\ref{eq:o-c2-3})] shows
finite--size scaling consistent with a phase transition in the Ising universality class,
in contrast with the triangular-lattice Heisenberg model~\cite{Miyashita1984}.
Results are taken from classical Monte Carlo simulations of
$\mathcal{H}^{\sf FFM}_\square$~[Eq.~(\ref{eq:Hex})] for the parameter set
${\bb J}_A$~[Eq.~(\ref{eq:def.JA})], in the absence of magnetic field ($h=0$).
}
\label{fig:JT-h=0}
\end{figure}

\section{Triangular--lattice physics on a square lattice}
\label{section:simulations.for.JA}

In Sec.~\ref{sec:mean.field.triangles}, we established a connection between the square--lattice
frustrated ferromagnet, $\mathcal{H}^{\sf FFM}_\square$~[Eq.~(\ref{eq:Hex})], and the Heisenberg
antiferromagnet on a triangular lattice, $\mathcal{H}^{\sf AF}_\triangle$~[Eq.~(\ref{eq:H.trianglar.AF})],
in the case where the ground state of $\mathcal{H}^{\sf FFM}_\square$ is a 1D--spiral with
wave vector
\begin{eqnarray}
{\bb Q}^{\sf 1D}_{\sf 3sub}
    &=& \left( \frac{2\pi}{3}, 0 \right)
    \; \text{or} \;
    \left( 0, \frac{2\pi}{3} \right)  \; , \nonumber
\end{eqnarray}
[cf. Eq.~(\ref{eq:def.Q.1D.triangle})], corresponding to three-sublattice ``stripe'' order.
An example of a three-sublattice stripe state, with finite magnetisation, is shown in
Fig.~\ref{fig:novel.phases}(a).
The momentum set of the ground state manifold
\begin{equation}
Q_\triangle = \{( \pm 2\pi/3, 0 ), \ \ ( 0, \pm 2\pi/3 ) \},
\label{eq:tri-lat-points}
\end{equation}
has four components, whereas that of the triangular antiferromagnet has only two.


In what follows, we use classical Monte Carlo simulation to explore the properties
of  $\mathcal{H}^{\sf FFM}_\square$ at finite temperature and magnetic field, considering a
parameter set~${\bb J}_A$~[Eq.~(\ref{eq:def.JA})].
The results of these simulations are summarised in the phase diagram Fig.~\ref{fig:JTphase}.
The similarities to the magnetic phase diagram of the triangular-lattice antiferromagnet  
$\mathcal{H}^{\sf AF}_\triangle$, cf. Fig.~1 of \linecite{Seabra2011a}, are striking. 
At first sight, the main difference is only that the ordering temperature scale is roughly 
double that for $\mathcal{H}^{\sf FFM}_\square$.
As in the case of the triangular-lattice antiferromagnet~\cite{Seabra2011a}, it will be instructive to compare
the phase diagram where a proper finite-size scaling has been performed [Fig.~\ref{fig:JTphase}],
to one extracted from simulations of a fixed cluster size $L=90$ [Fig.~\ref{fig:JTphaseL90}].


\begin{figure*}[t!]
\centering
\includegraphics[width=0.7\paperwidth]{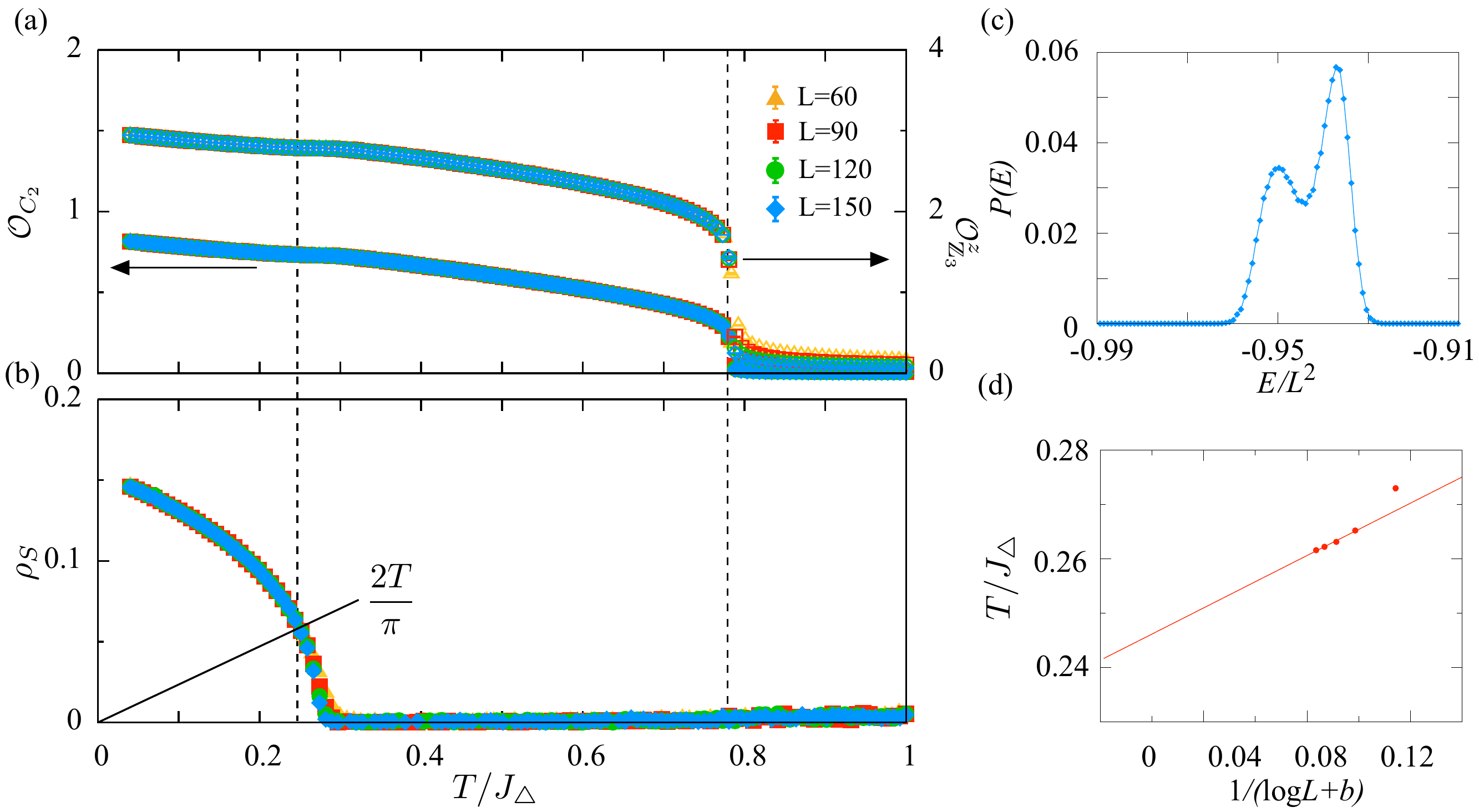}
\caption{%
(Color online).
Evidence for a collinear \mbox{$1/3$--magnetisation} plateau and coplanar ``Y--state'' in
a square--lattice frustrated ferromagnet.
(a) Temperature dependence of the order parameters
$\mathcal{O}_{C_2}$~[Eq.~(\ref{eq:o-c2-3})]
and $\mathcal{O}^z_{\mathds{Z}_3}$~[Eq.~(\ref{eq:o-c3-3})],
showing the onset of a collinear state with three-sublattice stripe order
($1/3$--magnetisation plateau), at $T \approx 0.78 J_\triangle$.
(b) Temperature dependence of the spin stiffness $\rho_S$~[Eq.~(\ref{eq:rhos-3})],
consistent with a BKT transition into the ``Y--state'' at \mbox{$T_L \approx 0.24 J_\triangle$}.
(c) Energy histogram calculated for $T = 0.776 J_\triangle$, showing the discontinuous
nature of the phase transition from paramagnet to $1/3$--magnetisation plateau.
(d) Finite--size scaling of the temperature associated with the BKT transition,
$T_L$~[Eq.~(\ref{eq:fss-bkt})].
Results are taken from classical Monte Carlo simulations of
$\mathcal{H}^{\sf FFM}_\square$~[Eq.~(\ref{eq:Hex})] for the parameter set
${\bb J}_A$~[Eq.~(\ref{eq:def.JA})], in applied magnetic field
$h/J_\triangle = 2$, cf.~dashed line in Fig.~\ref{fig:JTphaseL90}.
Temperature is measured in units of $J_\triangle$~[Eq.~(\ref{eq:def.J.triangle})].
\label{fig:JT-field}
}
\end{figure*}


While the two models have much in common, %
differences arise in the way in which ordered phases break lattice symmetries.
In both cases, under applied magnetic field, three-sublattice ordered phases break a discrete
$C_3 \cong \mathds{Z}_3$
symmetry, associated with the interchange of the different sublattices.
However the stripe--like order found in the square--lattice model also breaks
a $C_2$ lattice--rotation symmetry, when choosing between the two possible ordering
vectors, ${\bb Q}^{\sf 1D}_{\sf 3sub}$~[Eq.~(\ref{eq:def.Q.1D.triangle})].
This additional symmetry has a number of interesting consequences, described below.

\subsection{Ising transition at $h=0$}
\label{subsection:h0transition}

We consider first the limit of vanishing magnetic field, corresponding to
$h/J_\triangle=0$ in Fig.~\ref{fig:JTphase}.
In the absence of magnetic field, the Heisenberg antiferromagnet on a triangular
lattice $\mathcal{H}^{\sf AF}_\triangle$~[Eq.~(\ref{eq:H.trianglar.AF})]
has been argued to exhibit a finite--temperature transition,
linked with the proliferation of $\mathds{Z}_2$ vortices
associated with spin chirality~\cite{Miyashita1984}.
Whether this process corresponds to either a true phase transition or a crossover is still
under debate~\cite{Kawamura2010,Southern1995,Delamotte2010,Hasselmann2014}.


The situation in the square--lattice frustrated ferromagnet
$\mathcal{H}^{\sf FFM}_\square$~[Eq.~(\ref{eq:Hex})] is quite different.
At $h=0$, a clear phase transition is observed, associated with the breaking of
lattice--rotation symmetry by three--sublattice stripe order.
The relevant order parameter is
$\mathcal{O}_{C_2}$~[Eq.~(\ref{eq:o-c2-3})], and in Fig.~\ref{fig:JT-h=0}
we show a scaling plot of simulation results for
\begin{eqnarray}
\mathcal{O}_{C_2} = L^{-\beta/\nu}\mathcal{\tilde{O}}(tL^{1/\nu}),
 \label{eq:fss-c2}
\end{eqnarray}
as a function of the reduced temperature
\begin{eqnarray}
t = \frac{T_c-T}{T} \; ,
\end{eqnarray}
where the Ising critical exponents
\begin{eqnarray}
\nu = 1
\quad , \quad
\beta = \frac{1}{8} \;,
\end{eqnarray}
are found to describe the data well.


We do not observe behaviour explicitly related to the unbinding of ${\mathbb Z}_2$ vortices
in simulation of $\mathcal{H}^{\sf FFM}_\square$.
However, just as in simulations of $\mathcal{H}^{\sf AF}_\triangle$
[cf. Ref.~\onlinecite{Seabra2011a}], the correlation length associated with
transverse components of spin becomes very large as $h \to 0$, making it
difficult to draw definitive conclusions.
%

\subsection{Double transition for $0 \lesssim h \le 3 J_\triangle$}
\label{subsection:double_phase_transition}

We now consider the phases found for low to intermediate values of magnetic field in the phase diagram Fig.~\ref{fig:JTphase}.
Here, a double phase transition is observed as the system is cooled down
from the paramagnet  --- cf. Fig.~\ref{fig:JT-field}, for $0 \lesssim h/J_\triangle \le 3$.
As the system is cooled from the high--temperature paramagnet, it first enters the
collinear $1/3$--magnetisation plateau, as described by the rise of the order parameters
in Fig.~\ref{fig:JT-field}(a).
This process is a single phase transition, which simultaneously breaks the $C_2$
lattice--rotational symmetry  [Eq.~(\ref{eq:o-c2-3})], and the $\mathds{Z}_3$ translational
symmetry [Eq.~(\ref{eq:o-c3-3})] associated with the three-sublattice order.
For values of magnetic field \mbox{$1.5\lesssim h/J_\triangle < 3$}, a double distribution
in the internal energy is clearly observed at the transition temperature [Fig.~\ref{fig:JT-field}(c)],
which shows that the transition is of first order.
We cannot observe this behaviour for lower magnetic fields, presumably due to the
increased finite-size effects.


In the case of the Heisenberg antiferromagnet on a triangular lattice,
different $1/3$--magnetisation plateau states are connected by a 3--fold
permutation symmetry, and the phase transition into the paramagnet
is continuous~\cite{Seabra2011a}.
Meanwhile, in the case of the square--lattice frustrated ferromagnet, the
permutation of different sublattices is complemented by a lattice rotation,
enlarging the symmetry from $\mathds{Z}_3$ to $\mathds{Z}_2\times\mathds{Z}_3$,
and the corresponding phase transition is first order.
This is reminiscent of the 6-state Potts model in 2D, whose
ordering phase transition is known to be first order~\cite{Baxter1973,Baxter1978}.


The canted $Y$ state is found by lowering the temperature further from the plateau phase.
In addition of breaking the $C_2$ and $\mathds{Z}_3$ symmetries, this phase displays
algebraic order in the $S^x$--$S^y$ plane, as shown by the finite spin stiffness in Fig.~\ref{fig:JT-field}(b).
Assuming a Berezinski-Kosterlitz-Thouless (BKT) transition~\cite{Nelson1977}, the transition
temperature in a finite--size system is found via the jump in spin stiffness
\begin{eqnarray}
T_L = \frac{\pi}{2} \times \Delta\rho_S,
\label{eq:def.T.of.L}
\end{eqnarray}
which can then be finite-size scaled in a characteristic logarithmic fashion~\cite{Weber1988,Seabra2011a}
\begin{eqnarray}
T_L = T_{\sf BKT} \Big(1 + \frac{1}{2}\frac{1}{\log L + \log b}  \Big),
\label{eq:fss-bkt}
\end{eqnarray}
as shown in Fig.~\ref{fig:JT-field}(d).


%
The finite-size corrections to this transition are observed to be rather large.
From simulations of a fixed cluster size $L$, cf. the phase diagram for $L$$=$$90$
in Fig.~\ref{fig:JTphaseL90}, it is unclear if the BKT transition merges with the
$C_2\otimes\mathds{Z}_3$ transition for fields $h\lesssim 0.6$, i.e. if there is a
single transition from the paramagnet to the $Y$ state, or two.
However, these two phase transitions are observed to be clearly separated once the
proper finite-size scalings are performed, and no direct transition from the paramagnet
into the $Y$ phase can be reported for fields $h\ge0.4J_\triangle$ [Fig.~\ref{fig:JTphase}].
We have faded the region in Fig.~\ref{fig:JTphase} corresponding to values of field
$0<h\le0.4J_\triangle$, since finite-size effects become very large in this region and this analysis
becomes unreliable. However, the available data still suggests the presence of a double phase transition
as the field approaches $h\rightarrow 0$.
This behaviour very closely matches what is observed in the Heisenberg
antiferromagnet on a triangular lattice, which has been discussed in detail previously~\cite{Seabra2011a}.


For magnetic fields above the plateau $h>3J_\triangle$, we register a single
phase transition into the 2:1 canted state.
The location of several quantities, such as the jump in the spin stiffness, anomalies
in the $C_2$ and $\mathds{Z}_3$ order--parameter susceptibilities, and in the specific heat, coincide
within resolution, after they are properly finite-size scaled.
We do not observe any discontinuity in the internal energy as a function of temperature,
or a bimodal energy distribution at the transition temperature, in the clusters
simulated.
None the less, the maximum value of the specific heat scales roughly with
$N$, and the temperature at which it is found scales roughly with $1/N$,
behaviour consistent with a first-order transition.
%

\section{Consequences of an enlarged ground state manifold}
\label{section:simulations.for.JB}

\begin{figure}[t!]
\centering
\includegraphics[width=\columnwidth]{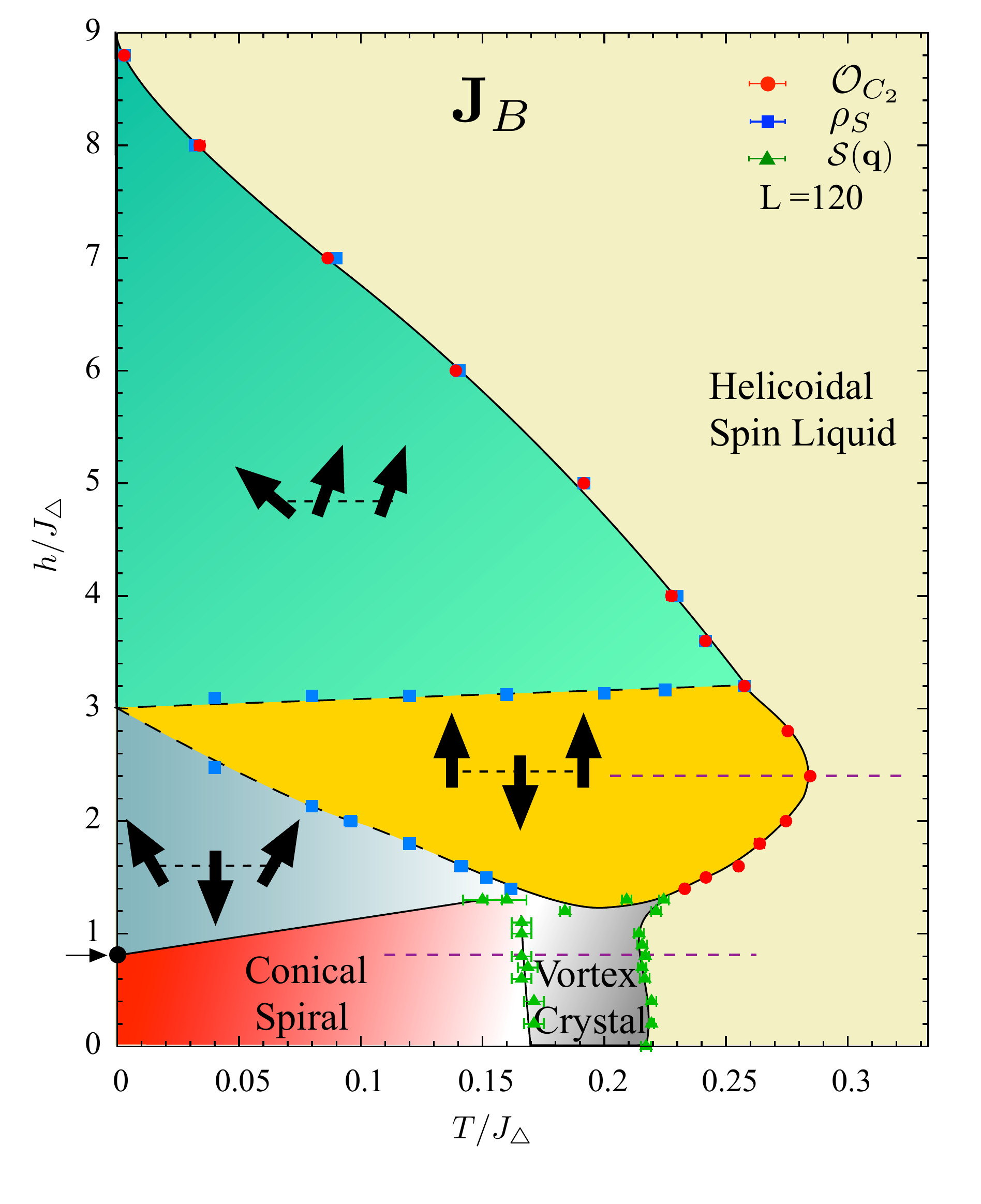}
\caption{
(Color online).
Finite--size phase diagram of a square--lattice frustrated ferromagnet in applied
magnetic field, showing a classical spin liquid and, at low fields, a vortex crystal
[cf.~Fig.~\ref{fig:novel.phases}(c)].
The other phases found are a conical spiral state interpolating to zero magnetisation;
a coplanar ``Y--state'' and a collinear \mbox{$1/3$--magnetisation} plateau
[cf.~Fig.~\ref{fig:novel.phases}(a)] at intermediate magnetisation; and a
coplanar 2:1 canted state interpolating to saturation.
Results are taken from classical Monte Carlo simulations of
$\mathcal{H}^{\sf FFM}_\square$~[Eq.~(\ref{eq:Hex})] for the parameter set
${\bb J}_B$~[Eq.~(\ref{eq:def.JB})], corresponding to a point of high degeneracy.
Phase boundaries were extracted from anomalies in the order parameter
associated with lattice rotations~[Eq.~(\ref{eq:o-c2-3})],
 spin stiffness $\rho_S$~[Eq.~(\ref{eq:def.S.perp})]
and the spin structure factor
${\mathcal S}({\bf q})$~[Eq.~(\ref{eq:S(q)-def})], for a cluster of
$N = L^2 = 120^2= 14400$ spins.
Temperature and magnetic field are measured in units of
$J_\triangle$~[Eq.~(\ref{eq:def.J.triangle})].
The horizontal dashed lines %
corresponds to the temperature cuts used in
Figs.~\ref{fig:ring-T-evolution}--\ref{fig:JRingIIC} and Fig.~\ref{fig:JAlowfield}.
\label{fig:JRphase}
}
\end{figure}

We now consider the properties of $\mathcal{H}^{\sf FFM}_\square$~[Eq.~(\ref{eq:Hex})]
for the parameter set ${\bb J}_B$~[Eq.~(\ref{eq:def.JB})],
as summarised in the phase diagram, Fig.~\ref{fig:JRphase}.
This parameter set corresponds to the point in the classical ground
state phase diagram [Fig.~\ref{fig:classical.ground.state.phase.diagram}],
where the line of 1D spirals with three-sublattice order terminates
on the boundary with the 2D spiral phase.
The behaviour of the model at this special point is determined by the
highly--degenerate manifold of states described in Sec.~\ref{subsec:ring-equations}.
And, as we shall see, the ``ring'' structure defined in Eq.~(\ref{eq:ring-qx-qy})
leaves a characteristic fingerprint in the spin structure factor, for all temperatures
and values of magnetic field.


For a generic choice of parameters on the boundary between 1D and 2D spirals,
it is not possible to find commensurate wave vectors $\bb Q$ which satisfy
Eq.~(\ref{eq:ring-qx-qy}).
However for specific choices of parameters $\bb J$, commensurate wave vectors do
exist, and the parameter set which offers the greatest number of commensurate
states is in fact set ${\bb J}_B$.
Here the wave vectors
\begin{eqnarray}
{\bb Q}_{R} = \{ (0,2\pi/3), (\pi/2,\pi/3), (\pi/5,3\pi/5)\},
\label{eq:QR}
 \end{eqnarray}
together with their reflections and mirrors, fulfill Eq.~(\ref{eq:ring-qx-qy}).
In total, this give 20 different commensurate wave vectors, compatible
with a wide range of cluster sizes.%


\begin{figure*}[t!]
\centering
\includegraphics[width=0.8\paperwidth]{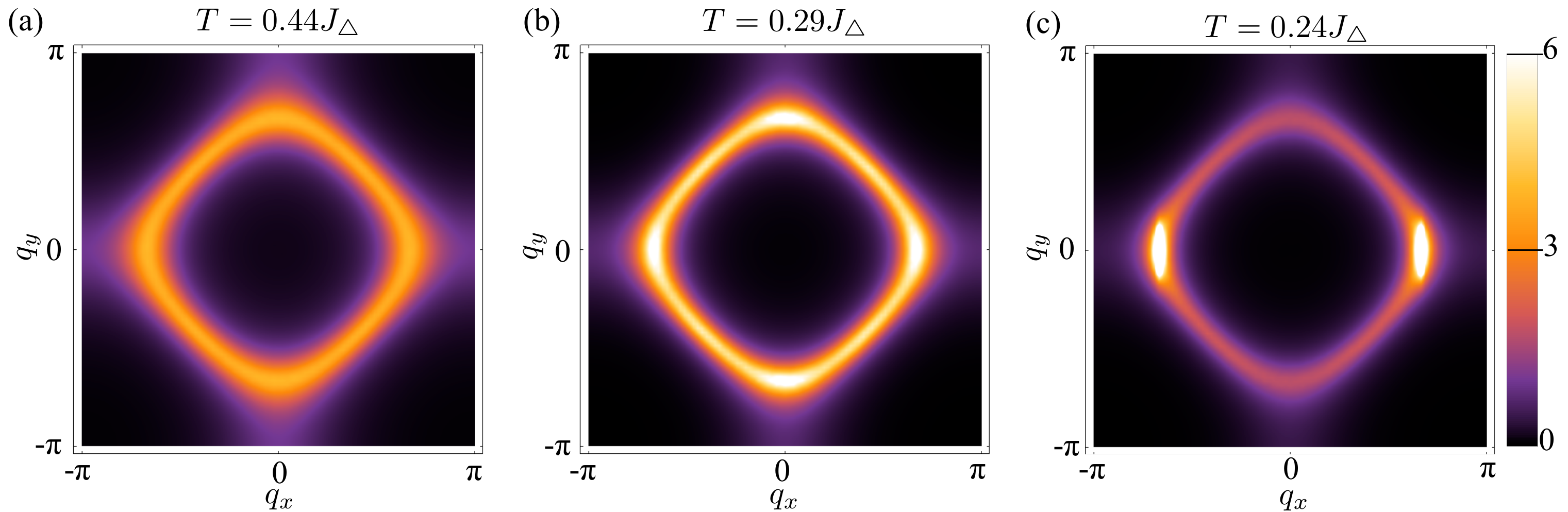}
\caption{%
(Color online).
Transition from the helicoidal spin liquid into a collinear $m=1/3$ magnetisation plateau,
in applied magnetic field, as revealed by the spin structure factor
$\mathcal{S}(\bold{q})$ [Eq.~(\ref{eq:S(q)-def})].
(a) Spin correlations for $T=0.44J_\triangle$, showing a ``ring'' feature characteristic
of the high--temperature spin liquid.
(b) Spin correlations for $T=0.29J_\triangle$, approaching the transition into an ordered state,
showing an enhancement of fluctuations near to potential ordering wave vectors.
(c) Spin correlations for $T=0.24J_\triangle$, within the collinear $m=1/3$ magnetisation plateau,
showing Bragg peaks at ${\bb Q}^{\sf 1D}_{\sf 3sub}$ [Eq.~(\ref{eq:def.Q.1D.triangle})], coexisting
with the ``ring'' feature.
Results are taken from classical Monte Carlo simulations of
$\mathcal{H}^{\sf FFM}_\square$~[Eq.~(\ref{eq:Hex})], for a cluster of $L^2=90^2$ spins,
with exchange parameters ${\bb J}_B$~[Eq.~(\ref{eq:def.JB})].
Magnetic field was set to $h = 2.4 J_\triangle$, corresponding to the dashed in line
in Fig.~\ref{fig:JRphase}.
The ${\bb q} = 0$ component of the spin correlations has been subtracted for clarity.
}
\label{fig:ring-T-evolution}
\end{figure*}


We begin by establishing the finite--temperature phase diagram, Fig.~\ref{fig:JRphase},
using  Monte Carlo simulations for a cluster of size  $L=120$.
For values of applied field $h\gtrsim1.2J_\triangle$, the phase diagram closely follows that
of the  parameter set ${\bb J}_A$~[Eq.~(\ref{eq:def.JA})] (and therefore the Heisenberg
antiferromagnet on a triangular lattice,
$\mathcal{H}^{\sf AF}_\triangle$~[Eq.~(\ref{eq:H.trianglar.AF})]), displaying the succession of canted
$Y$ phase, collinear $1/3$--magnetisation plateau and canted $2:1$ phase as a function of increasing magnetic field.


Simulations become very challenging at lower values of applied field, $h\lesssim1.2J_\triangle$,
making it very difficult to extrapolate results to the thermodynamic limit.
Still, in this region we can establish the presence of two additional non-coplanar phases.
At very low temperatures, a conical spiral phase described by a single wave-vector is favoured by
thermal fluctuations, as described in Section~\ref{subsection:low-field-phase-diag-jt}.
At intermediate temperatures, between this phase and the disordered paramagnet phase,
a novel conical phase, with a more complex spin texture, is stabilised, as described in 
Section~\ref{subsection:order1Q4Q} --- cf. Fig.~\ref{fig:novel.phases}(c).


\begin{figure*}
\centering
\includegraphics[width=0.7\paperwidth]{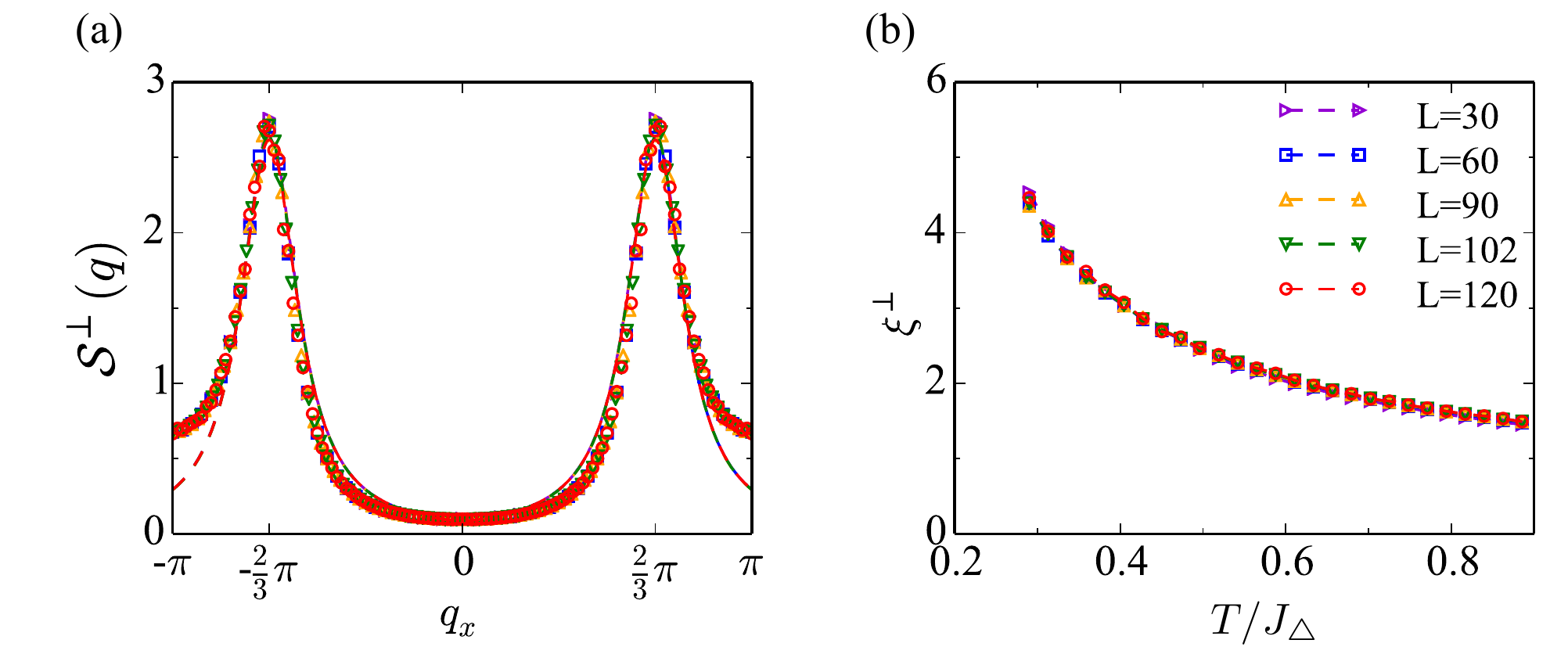}
\caption{%
(Color online)
Characterisation of the helicoidal spin liquid in applied magnetic field $h=2.4J_\triangle$.
(a) Structure factor $\mathcal{S}^\perp(q_x,0)$~[Eq.~(\ref{eq:S.perpendicular.of.q})],
showing a cross section of the characteristic ``ring'' of scattering
[cf.~Fig.~\ref{fig:ring-T-evolution}(a)].
Data for a wide range of system sizes collapses onto a single curve.
Dashed lines are fits to a Lorentzian [Eq.~(\ref{eq:lorentzian})],
from which the correlation length $\xi$ is obtained.
(b) Temperature--dependence of the correlation length $\xi$.
Results are taken from classical Monte Carlo simulations of
$\mathcal{H}^{\sf FFM}_\square$~[Eq.~(\ref{eq:Hex})],
for clusters of $N = L \times L$ spins, with $T=0.45J_\triangle$
and the same values of exchange and magnetic field as Fig.~\ref{fig:ring-T-evolution}.
\label{fig:JRingIIA}
}
\end{figure*}


All the ordered phases at ${\bb J}_B$ are described by wave vectors belonging to,
or combinations of, ${\bb Q}_{R}$ [Eq.~(\ref{eq:QR})].
All the transitions from the disordered phase are very clearly observed to be of first order.
The overall  ordering-temperature scale is strongly reduced from the  ${\bb J}_A$ case,
both in units of $J_\triangle$ and $|J_1|$, indicating the larger role played by frustration
in this case, suppressing the ordering tendency of the model.

\subsection{Spin-liquid phase from a ring of correlations}
\label{section:ring}

We consider first the nature of the paramagnetic phase found
for all values of magnetic field at sufficiently high temperature,
as shown in the phase diagram Fig.~\ref{fig:JRphase}.
The highly--degenerate ground--state manifold of $\mathcal{H}^{\sf FFM}_\square$
at the \mbox{parameter set ${\bb J}_B$~[Eq.~(\ref{eq:def.JB})]} is not just of concern
at $T=0$, but also has profound consequences at finite temperature.
Its effects are most obvious in the spin structure factor
$\mathcal{S}(\bf q)$~[Eq.~(\ref{eq:S(q)-def})] which takes on a finite value
for all wave vectors $\bb q$ which satisfy, or nearly satisfy, the ``ring''
condition Eq.~(\ref{eq:def.JB}).
A corresponding ring structure can be seen in $\mathcal{S}(\bf q)$ for {\it all}
of the ordered phases shown in Fig.~\ref{fig:JRphase}, and is equally prominent
for $h=0$ [Fig.~\ref{fig:novel.phases}(b)], and for $h=2.4J_\triangle$
[Fig.~\ref{fig:ring-T-evolution} .
In the paramagnetic regions of the phase diagram, Fig.~\ref{fig:JRphase}, 
where no symmetries are broken, the extra degeneracy gives rise to a
classical spin liquid.


The nature of the correlations along the ring does not change in any fundamental way
when magnetic field is applied.
In Fig.~\ref{fig:ring-T-evolution} we show the evolution of the full structure factor 
$\mathcal{S}(\bf q)$ at \mbox{$h=2.4J_\triangle$}, as the temperature is lowered from 
the disordered phase into the $1/3$--magnetisation plateau.
At high temperatures and away from the transition [Fig.~\ref{fig:ring-T-evolution}(a)], the height
of $\mathcal{S}(\bf q)$ is very uniform, such that many wave vectors belonging to the ring, or
very close to it, contribute equally.
If the temperature is raised from this point, the height of $\mathcal{S}(\bf q)$ just decreases smoothly, and no crossover
into a standard paramagnet is observed at any temperature scale.
When the system is cooled down to near the phase transition $T\approx0.29J_\Delta$, Fig.~\ref{fig:ring-T-evolution}(b),
peaks begin to develop at the wave vectors $Q_\triangle$ [Eq.~(\ref{eq:tri-lat-points})],
corresponding to incipient three-sublattice magnetic order.
Inside the  magnetisation plateau at even lower temperature, Fig.~\ref{fig:ring-T-evolution}(c), Bragg peaks (from the
$\mathcal{S}^{z}({\bb q})$ component of the structure factor) are observed, associated with
the broken $\mathds{Z}_3$ symmetry discussed in Sec.~\ref{section:simulations.for.JA}.
A diffuse ring of low-lying excitations can still be observed in this phase, attesting to the
pervasiveness of the ring correlations.


In order to further clarify the physical implications of this ring in the structure factor, we now focus on the transverse component $\mathcal{S}^\perp(\mathbf{q})$ [Eq.~(\ref{eq:S.perpendicular.of.q})].
Fig.~\ref{fig:JRingIIA}(a) shows a cut of $\mathcal{S}^\perp(\mathbf{q})$ along the line $q_y=0$
in the disordered phase for $h=2.4J_\triangle$ and $T=0.45J_\triangle$.
We find that the structure factor can be fitted quite well to a (double) Lorentzian expression
of the form
\begin{eqnarray}
\mathcal{S}^\perp(q) =\frac{A^2}{\xi^{-2}+(q-q_0)^2},
\label{eq:lorentzian}
\end{eqnarray}
centred at $q_0 = \pm 2\pi/3 $. Similar fits can be performed for different cuts of
$\mathcal{S}^\perp(\mathbf{q})$ in the $Q_x-Q_y$ plane.


The width of the structure factor around $q_0$ is controlled by the correlation length $\xi$,
which sets a finite length-scale for correlations.
$\mathcal{S}^\perp(q)$  shows practically no finite-size dependence for the clusters studied,
attesting to the short-range nature of the correlations.
The correlation length $\xi$ decays slowly with temperature [Fig.~\ref{fig:JRingIIA}(b)],
which indicates the stability of the ring correlations for a wide range of temperatures.


Our next target is the {\it total} structure factor arising
from low-energy excitations associated with the ring.
First, we need identify the wave vectors $\bb Q_{\sf ring}$  contributing
to the total ring structure factor.
A finite $\xi$ means that there are momenta outside the $T$$=$$0$ ring
[Eq.~(\ref{eq:ring-qx-qy})] which will  contribute to the total structure factor,
[cf. Fig.~\ref{fig:novel.phases}(b)].
In order to capture these wave vectors with a finite spectral weight,
we  take the one-sublattice spin-wave dispersion Eq.~(\ref{eq:dispersion1SL})
 [see Appendix~\ref{section:Appendix_spin_wave}]
and impose an energy cutoff of $\omega_c=0.01J_\triangle$ on it.
Alternatively, using the inverse of the correlation length of Fig.\ref{fig:JRingIIA}(b) as input,
we can predict approximately the wave-vectors which will be thermally activated.
Both approaches provide the same set of wave vectors
 belonging to the enlarged ring,
but the latter, correlation-length based method becomes unwieldy and computationally
more expensive for large system sizes.

\begin{figure}[b!]
\centering
\includegraphics[width=0.7\columnwidth]{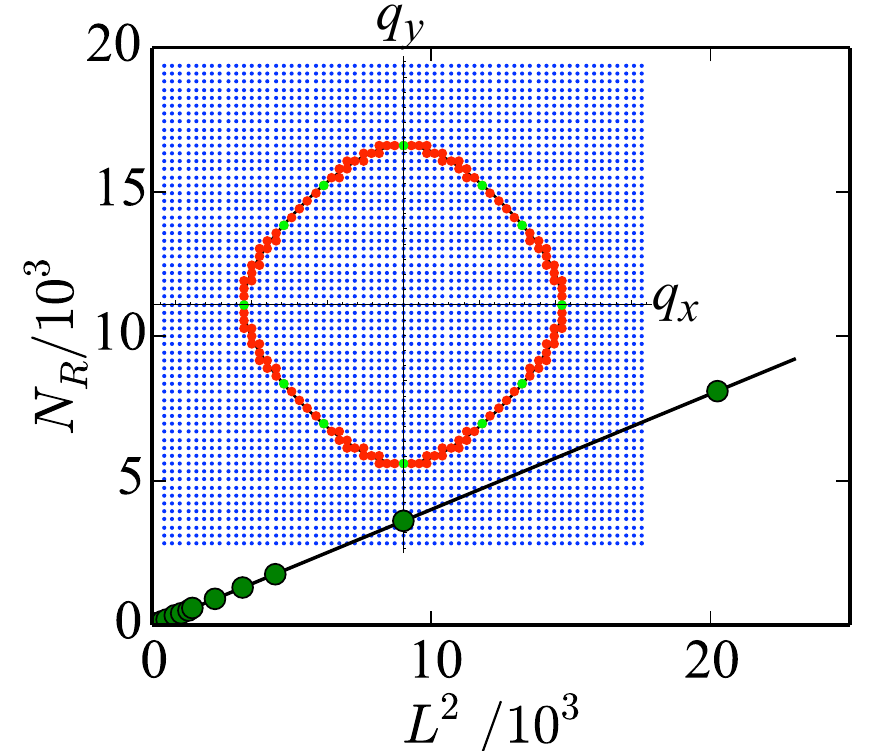}
\caption{
(Color online)
Evidence that the number of spin configurations contributing to the
helicoidal spin liquid, $N_R$, scales linearly with the system--size,
$N = L^2$.
$N_R$ was estimated on the basis of the number of states lying close
to the ring defined by [Eq.~(\ref{eq:ring-qx-qy})], for
parameters ${\bb J}_B$~[Eq.~(\ref{eq:def.JB})] and system sizes
ranging from $L = 12^2$ to $L=450^2$, as described in Section~\ref{section:ring}.
Inset : wave vectors of states contributing to $N_R$ for $L=60$.
\label{fig:JRing-Nring}
}
\end{figure}

The number of momentum vectors, $N_R$, thermally activated is found to scale
$linearly$ with the cluster size $N_R\sim L^2$, for a very wide range of system sizes
[Fig.~\ref{fig:JRing-Nring}].
The area covered by the ring is given roughly by the product of its width with its perimeter, which
scales with $L$.


This implies that there are approximately $L$ cuts across the ring such as
Fig.~\ref{fig:JRingIIA}(a).
The ring has a fixed width in momentum space, set by $\xi^{-1}$, which is basically
independent of $L$, see Fig.~\ref{fig:JRingIIA}(b).
Since the resolution in momentum space scales with system size as $L^{-1}$, the 
number of points allowed inside each cut of the form Fig.~\ref{fig:JRingIIA}(a) 
must scale as $L$.
Therefore, the number of points covered by the ring scales as \mbox{$N_R\sim L\times L$},
and is an $extensive$ property of the system.

\begin{figure}[t]
\centering
\includegraphics[width=0.8\columnwidth]{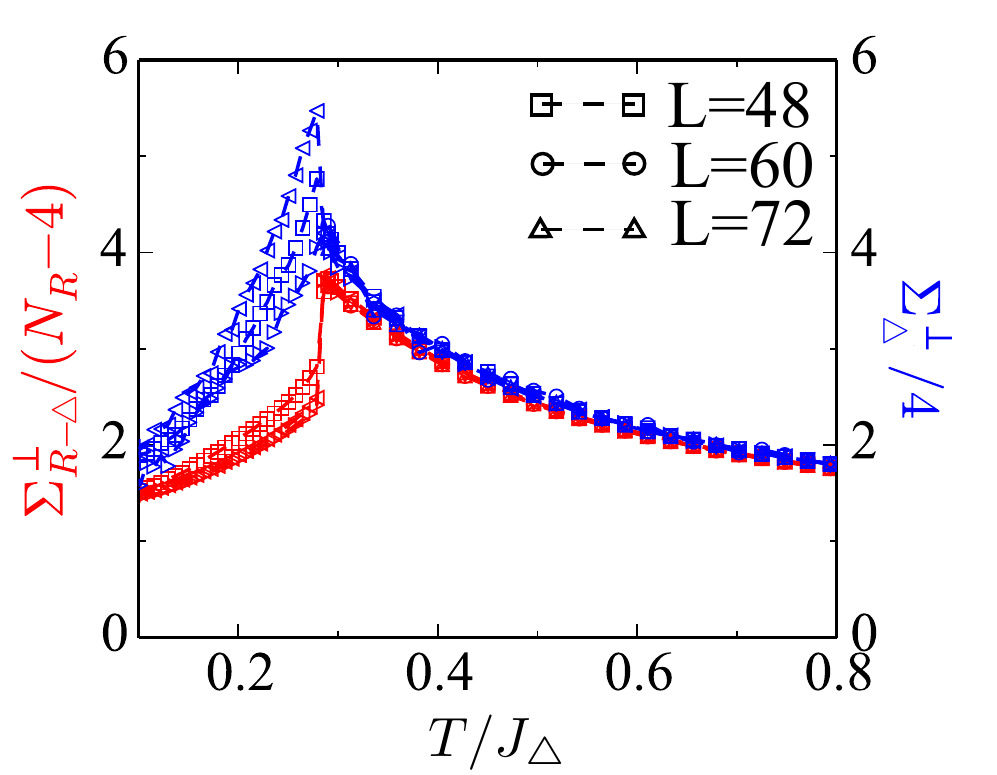}
\caption{
(Color online).
Comparison between the spectral weight associated with the ordering vectors
of the 1/3--magnetisation plateau, and that associated with the remaining 
wave--vectors belonging to the ring [cf. Fig.~\ref{fig:ring-T-evolution}].
The total spectral weight in the ring, $\Sigma^\perp_{\triangle}$ [Eq.~(\ref{SR})],
undergoes a discontinuous drop at the ordering temperature
$T = 2.8 J_\triangle$, but remains  finite in the ordered state.
Fluctuations at the ordering vector, $\Sigma^\perp_{R-\triangle}$ [Eq.~(\ref{SRT})],
are associated with a discontinuous rise at the ordering temperature.
Results are taken from classical Monte Carlo simulations of
$\mathcal{H}^{\sf FFM}_\square$~[Eq.~(\ref{eq:Hex})] for
system sizes $L = 48,\, 60,\, 70$, and parameters
${\bb J}_B$~[Eq.~(\ref{eq:def.JB})] with
$h = 2.4 J_\triangle$.  %
\label{fig:JRing-uniform}
}
\end{figure}

In order to gain more insight into this state, we define an integral of the
transverse structure factor $S^\perp ({\bf q})$ [Eq.~(\ref{eq:S.perpendicular.of.q})]
over the manifold of states contributing to the ring
\begin{align}
\label{SR}
	\Sigma^\perp_R
	& =  \sum_{\mathbf{q}} \mathcal{S}^\perp(\mathbf{q}) \quad \forall \quad \mathbf{q}
		\in \{\mathbf{q}|\omega_{\sf 1SL}({\bf q}) < \omega_c\} \; ,
\end{align}
as well as a sum over the discrete wave vectors $Q_\triangle$ [Eq.~(\ref{eq:tri-lat-points})]
associated with three-sublattice order
\begin{align}
\label{ST}
	\Sigma^\perp_\triangle
	& = \sum_{\mathbf{q} \in \{ Q_\triangle \}} \mathcal{S}^\perp(\mathbf{q})
		 \; ,
\end{align}
and the difference of the two
\begin{align}
\label{SRT}
	\Sigma^\perp_{R-\triangle}
	& = \Sigma^\perp_R-\Sigma^\perp_\triangle \; .
\end{align}
Fig.~\ref{fig:JRing-uniform} shows how these measures of the ring, normalised
to the number of contributing wave vectors,
evolve across the transition from the high temperature spin--liquid
into the $1/3$--magnetisation plateau for $h=2.4J_\triangle$.
It is clear that the $\mathcal{S}(\bb q)$ ring has a uniform height at
temperatures away from the phase transition.
Right above the transition, enhanced fluctuations in $S^x-S^y$ herald three-sublattice order, but the defining
characteristics of the ring survive both in the disordered and ordered limits.


\begin{figure}[t]
\centering
\includegraphics[width=\columnwidth]{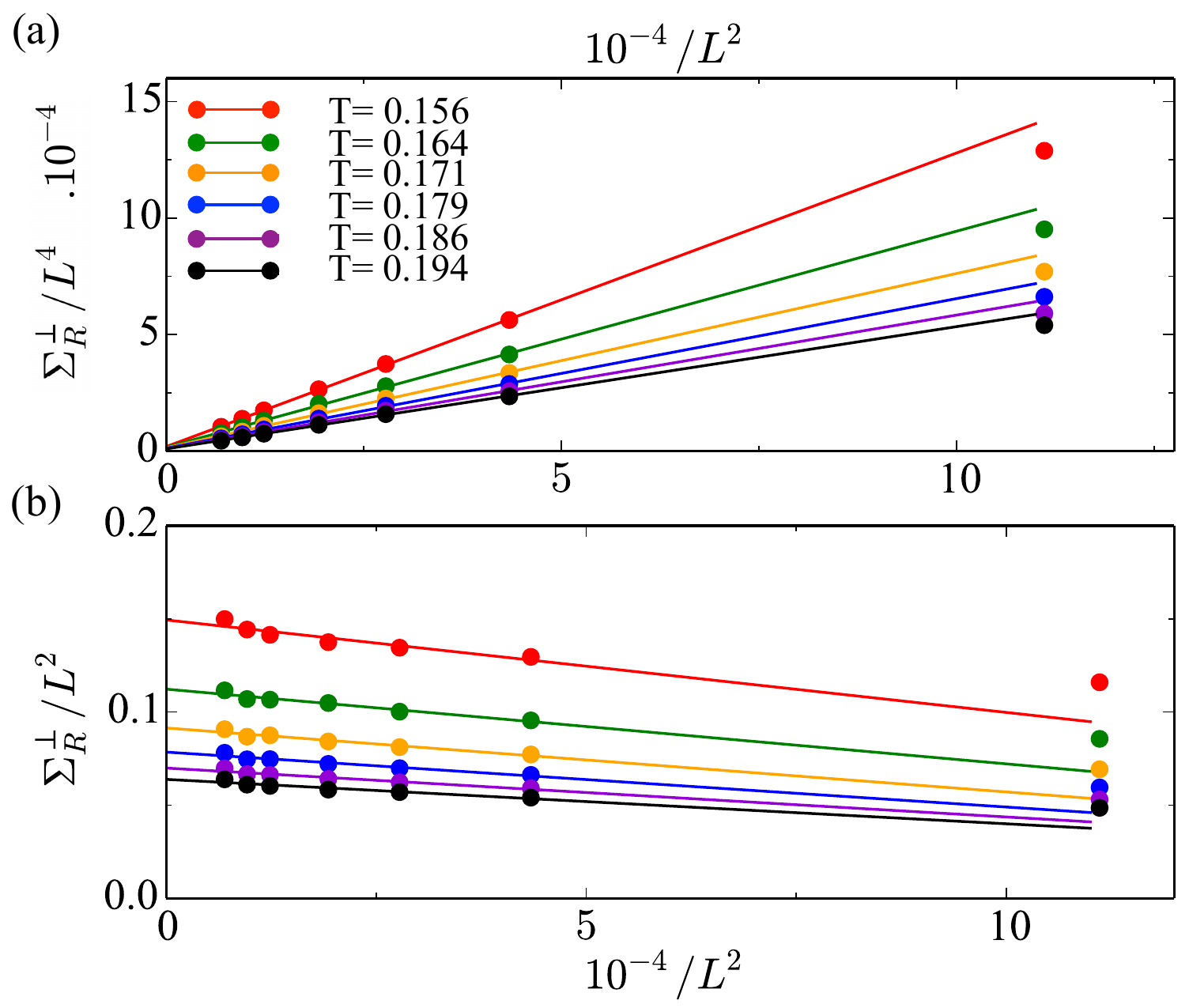}
\caption{%
(Color online).
Evidence for spin--liquid behaviour in the finite--size scaling of the integrated
structure factor, $\Sigma^\perp_R$~[Eq.~(\ref{SR})].
(a) The total weight associated with spin configurations contributing
to the ``ring'' vanishes as $1/L^2$ in the thermodynamic limit.
(b) The  structure factor per spin summed over all ring points, $\Sigma^\perp_R/L^2$,
scales linearly with the system size and thus persists in the thermodynamic limit.
Lines are a linear fit to the data for the largest system sizes.
Results are taken from classical Monte Carlo simulations of
$\mathcal{H}^{\sf FFM}_\square$~[Eq.~(\ref{eq:Hex})]
for $h=2.4J_\triangle$ and $T=0.45J_\triangle$,
with exchange parameters ${\bb J}_B$~[Eq.~(\ref{eq:def.JB})].
\label{fig:JRingIIC}
}
\end{figure}

The structure factor per spin 
\begin{eqnarray}
\left(\Sigma_R/L^2 \right)/ N_R \sim \Sigma_R/L^4
\end{eqnarray}
evaluated at a {\it typical} wave vector ${\bb Q}_{\sf ring}$ [Eq.~(\ref{eq:ring-qx-qy})], 
is itself not an extensive quantity.
Instead it vanishes as $1/L^2$, as can be seen in Fig.~\ref{fig:JRingIIC}(a) for a variety
of temperatures  in the disordered region.
This is also the behaviour observed in a standard paramagnetic region.
However, the {\it total} ring structure factor per spin $\Sigma_R/L^2$ is a function of
$N_R \sim L^2$ and should therefore be an {\it extensive} quantity.
Fig.~\ref{fig:JRingIIC}(b) shows that $\Sigma_R/L^2$ converges to a finite
value as $L^{-2}\rightarrow0$, and therefore $\Sigma_R$ is a non-zero
quantity in the thermodynamic limit.
This is a direct demonstration of an extensive, non-trivial degeneracy
in the thermodynamic limit, an hallmark of a spin liquid.


\subsection{Low--temperature conical spiral}
\label{subsection:low-field-phase-diag-jt}

We now turn to the nature of the low--temperature ordered state
found for $h/J_\triangle \lesssim 1$ and $T/J_\triangle \lesssim 1.7$
in the phase diagram Fig.~\ref{fig:JRphase}.
We find that specific ordered states, with a wave vector belonging to the
ring--manifold, Eq.~\ref{eq:ring-qx-qy}, are selected at low temperature.
None the less, ``ring'' seen in the spin structure factor $\mathcal{S}(\bf q)$~[Eq.~(\ref{eq:S(q)-def})]
survives in these ordered phases, as shown in Fig.~\ref{fig:JRing-uniform}.
For magnetic fields \mbox{$h\gtrsim1.2J_\triangle$}, the magnetisation process
of the triangular lattice antiferromagnet is recovered,  %
as described in Section \ref{section:simulations.for.JA}
for the parameter set ${\bb J}_A$~[Eq.~(\ref{eq:def.JA})].
For values of field $ 0 \le h\lesssim1.2J_\triangle$, we observe a strong competition
between the coplanar $Y$ state and different conical states, all with uniform magnetisation $S^z_i$.


\begin{figure*}[t]
\centering
\includegraphics[width=0.65\paperwidth]{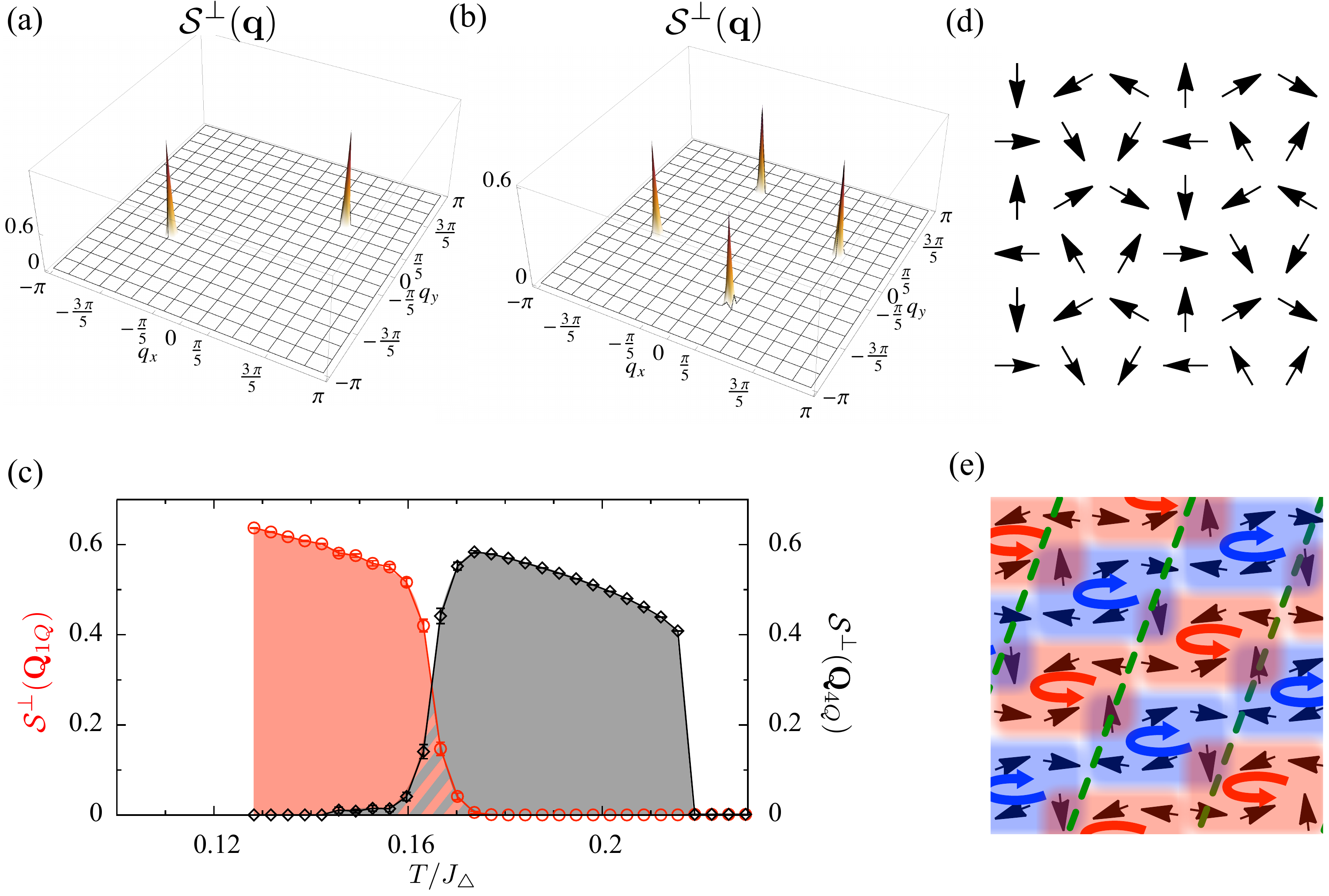}
\caption{
(Color online).
Finite--temperature phase transition between conical spiral state
and magnetic vortex crystal found for low value of magnetic field 
in the phase diagram Fig.~\ref{fig:JRphase}.
(a)~Structure factor $\mathcal{S}^\perp({\bb q})$~[Eq.~(\ref{eq:S.perpendicular.of.q})],
within the conical--spiral state for $T=0.13\  J_\triangle$,
showing peaks at $\mathbf{Q}_{1Q}$~[Eq.~(\ref{eq:Q_1Q})].
(b)~Structure factor within the magnetic vortex crystal for $T=0.175\  J_\triangle$,
showing Bragg peaks at $\mathbf{Q}_{4Q}$~[Eq.~(\ref{eq:Q_4Q})].
(c)~Temperature--dependence of the peaks, showing
a phase transition from conical spiral to magnetic vortex crystal
at $T \approx 1.6 \  J_\triangle$ [cf. Fig~\ref{fig:JRphase}].
(d)~Illustrative spin configuration in the plane perpendicular to magnetic field associated
with the $1Q$ conical--spiral state.
(e)~Illustrative spin configuration in the plane perpendicular to magnetic field associated
with the $4Q$ magnetic vortex crystal.
Results for $\mathcal{S}^\perp({\bb q})$ are taken from classical Monte Carlo simulations of
$\mathcal{H}^{\sf FFM}_\square$~[Eq.~(\ref{eq:Hex})] for $h=2.4 J_\triangle$,
with exchange parameters ${\bb J}_B$~[Eq.~(\ref{eq:def.JB})].
\label{fig:JAlowfield}
}
\end{figure*}


In order to better understand this behaviour, we have carried out a low-temperature
expansion of the spin-wave excitations around the $T=0$ classical ground--state manifold, 
which allows the calculation of corrections to the free energy as a power series in $T$.  
Details of these calculations are given in Appendix~\ref{subsection:Entropy_harmonic_approximation}.
In the absence of magnetic field, this expansion
predicts that low-temperature fluctuations favour a
coplanar spiral state described by a single incommensurate
wave vector very close to
\begin{eqnarray}
{\bf Q}_{1Q} = (\pi/3,\pi/2)
\label{eq:Q_1Q}
\end{eqnarray}
(plus associated mirrors and reflections), cf.~Fig~\ref{fig:CSW-I}.


When a small magnetic field is applied, the state favoured by fluctuations in the harmonic
approach is a {\it conical}, non-coplanar version of the $h=0$ state, cf. Fig.\ref{fig:JAlowfield}(a) and (d).
The wave-vector of the $h=0$ spiral is preserved in the $S^x-S^y$ plane, while all spins have a
constant magnetisation along the $S^z$ spin direction.
This state has been identified previously as the conical umbrella state in the anisotropic
triangular lattice~\cite{Veillette2005,alicea09,Okubo2012,Starykh2014}.
According to the low-temperature expansion, the entropy of the $Y$ coplanar state
increases with increasing value of magnetic field, while that of the conical state decreases.
For fields $h\gtrsim0.8J_\triangle$, the $Y$ state is predicted to be the preferred low-temperature state.


Monte Carlo simulations confirm that, for values of field $0\le h\lesssim 0.8 J_\triangle$, the
conical state is stabilised at low temperature, while for $0.8 \lesssim h\le 3 J_\triangle$
the $Y$ state is favoured.
The first-order phase transition between both phases is very difficult to observe in MC simulations, since
it happens after \emph{two} other symmetry-breaking phase transitions at higher temperature.
We also find, from e.g. a negative spin stiffness (not shown), that simulations seeded
with a conical spin texture with an in-plane wave vector ${\bf Q}_{1Q}=(\pi/3,\pi/2)$
show a strong tendency to become slightly incommensurate in the $S^x$$-$$S^y$ plane,
in full agreement with the harmonic approximation.
This selection of an incommensurate, non-coplanar phase by thermal fluctuations is an interesting
counterexample to the rule of thumb that fluctuations prefer collinear, or at worst coplanar, phases,
as a manifestation of the celebrated order-due-to-disorder effect, cf. \linecite{Henley1989}.

\subsection{Magnetic vortex crystal}
\label{subsection:order1Q4Q}

We next consider the nature of the ordered state found for 
intermediate temperatures, $1.7 < T/J_\triangle \lesssim 2.2$,
and low values of magnetic field, $h/J_\triangle \lesssim 1.2$, 
between the low--temperature conical spiral and the
high--temperature spin--liquid phase, as shown in the phase diagram Fig.~\ref{fig:JRphase}.
In this parameter range, Monte Carlo simulations started from a random
initial configuration often get trapped in local free-energy minima, resulting in
domain walls between different phases.
This hints at the presence of competing phases, and a first-order transition, 
at a lower temperature than the initial ordering transition.
Further simulations with open boundary conditions~\cite{luis-unpub}
reveal a tendency for the edge spins to be collinear with $S^z$, while the
bulk reproduces the same behaviour as for periodic boundary conditions.
We resort to comparing different parallel-tempering simulations for each value of field,
initialised from a variety of different ordered states, including random configurations,
mixing different states in a single replica, or across temperature space.
This approach does not permit a very accurate location of the phase transition, 
but gives us confidence on the phases present and on the overall topology of the phase diagram.


Carrying out this analysis, we find a constant-magnetisation state with a spin texture in the $S^x$--$S^y$
plane described by {\it four} wave vectors, cf.~Fig.~\ref{fig:novel.phases}(c).
Usually, multiple-Q states violate the fixed spin-length constraint $|\mathbf{S}_i|=1$,
and therefore are not favoured at low temperatures.
However, it is possible to construct a very specific $4Q$ state out of spirals
with the wave vector
\begin{eqnarray}
\mathbf{Q}_{4Q}=(3\pi/5,\pi/5)
\label{eq:Q_4Q}
\end{eqnarray}
and symmetry--related vectors.
%
%
For a given magnetisation 
\begin{eqnarray}
m = S_i^z = \text{const.} \; ,
\end{eqnarray}
the spin configuration associated with this $4Q$ state is
\begin{eqnarray}
S^x_i &=&\Big[1-m^2\Big]^{\frac{1}{2}}{\sf Re}[F({\bb r}_i)] \; ,\\
S^y_i &=&\Big[1-m^2\Big]^{\frac{1}{2}}{\sf Im}[F({\bb r}_i)] \; ,
\label{eq:4q-structure}
\end{eqnarray}
where
\begin{eqnarray}
F({\bb r}_i)&& =\ee{i\mathbf{Q}_A. \mathbf{r}_i}
+ \ee{i2\pi/5}\ee{-i.\mathbf{Q}_A. \mathbf{r}_i} \nonumber\\
&&+  \ee{i4\pi/15}\ee{i.\mathbf{Q}_B. \mathbf{r}_i}
+  \ee{-i8\pi/15}\ee{-i\mathbf{Q}_B. \mathbf{r}_i} \; ,
\end{eqnarray}
and
\begin{eqnarray}
\mathbf{Q}_A & = & (\pi/5, -3 \pi/5) \; , \nonumber\\
\mathbf{Q}_B & = & (3 \pi/5, \pi/5) \; .
\end{eqnarray}
For vanishing magnetisation ($m = S^z_i \equiv 0$), the $4Q$ state has the form
of a lattice of vortices shown in Fig.~\ref{fig:JAlowfield}(e).
By construction, the $4Q$ state interpolates smoothly to finite magnetisation,
where it takes on the form illustrated in Fig.~\ref{fig:novel.phases}(c).
The associated spin structure factor has four peaks
[cf.~Fig.~\ref{fig:JAlowfield}(b)].


Monte Carlo simulations  for a range of fields \mbox{$ 0 \lesssim h \lesssim 1.2 J_\triangle$},
 initialised with half of the replicas in the $1Q$ state, and the other
half in the $4Q$ state, paint a consistent picture of a
double phase transition, as shown in Fig.~\ref{fig:JAlowfield}(c), for $h=0.8J_\triangle$.
As the system is cooled down from the paramagnet, a region is found where the $4Q$
state is stabilised, followed by the $1Q$ state at a lower temperature.
We conclude that the inner phase transition is probably weakly first--order, since we observe 
no discontinuities in the internal energy, nor any significant release of entropy.
Unlike simulations of the $1Q$ state initialised with the ${\bf Q}_{1Q}=(\pi/3,\pi/2)$ wave vector,
the $4Q$ state does not show a tendency to become incommensurate, which attests to its stability.
We observe this scenario of two ordered phases down to the $h=0$ limit, where both phases lose their conical character by becoming coplanar.


The spin textures of the two non-coplanar states are fundamentally different.
A spiral phase described by a single wave vector breaks a reflection symmetry, which can
be detected, for example, by defining a chiral order parameter on the x-bonds
\begin{eqnarray}
\kappa_{\sf total}=\frac{1}{N}\sum_{i} \left(\mathbf{S}_i \times \mathbf{S}_{i+\hat{\mathbf{x}}} \right)^z.
\end{eqnarray}
Conversely, multiple-Q states typically have vanishing net chirality $\langle \kappa_{\sf total}\rangle=0$.
The $4Q$ state is no exception, and the chirality defined on each plaquette points along
the $S^z$ direction but with alternating signs, so that the total chirality is zero.
This describes a spin texture where the spin texture winds up in alternating directions
across the lattice, which may be visualised as a crystalline structure of alternating magnetic vortices
[Fig.~\ref{fig:JAlowfield}(e)].
We have not found other possible multiple-Q states with uniform $S^z$ which could be
stabilised in this model.

\section{Discussion of results}
\label{section:discussion}

In this Article, we have explored some of the novel phases which can arise in
a square--lattice frustrated ferromagnet in applied magnetic field.
In the process, we have seen a triangular---lattice antiferromagnet reborn on a
square lattice, and uncovered a spin--liquid and a vortex crystal, both with
finite magnetisation.
In what follows, we examine how these phases compare with known results
for frustrated antiferromagnets, and sketch some of
the possible consequences of quantum fluctuations, previously
touched upon in Refs.~\onlinecite{Sindzingre2009,Sindzingre2010}.

\subsection{Three-sublattice physics and $1/3$--magnetisation plateau}
\label{discussion_three-sublattice_physics}

The phase diagram for the square--lattice frustrated ferromagnet
$\mathcal{H}^{\sf FFM}_\square$~[Eq.~(\ref{eq:Hex})],
for the parameter set ${\bb J}_A$~[Eq.~(\ref{eq:def.JA})],
shown in Fig.~\ref{fig:JTphase}, is markedly different from what might
normally be expected in a square--lattice
frustrated magnet~\cite{zhitomirsky00}.
However, it is remarkably similar to that of the triangular--lattice
antiferromagnet, developed in \linecite{Seabra2011a}.
This similarity extends beyond the phases found ---  canted ``Y'' state,
collinear $1/3$--magnetisation plateau, and $2:1$ canted phase ---
and the topology of the resulting phase diagram, into the finite--size scaling
of the associated phase transitions (compare Fig.~\ref{fig:JTphase}
and Fig.~\ref{fig:JTphaseL90} of this Article with Fig.~1 and Fig.~2
of \linecite{Seabra2011a}).


These results are striking, since the collinear $1/3$--magnetisation plateau
is usually thought to be a hallmark of triangular--lattice physics, stabilised
by a delicate order--by--disorder effect \cite{Kawamura1985,Henley1989,Chubukov_Golosov_1991}.
Here, however, a robust $1/3$--magnetisation plateau, with three-sublattice
order, arises in a model more naturally associated with incommensurate phases.
In fact, for this particular parameter set, the only significant difference between
the square--lattice and triangular lattice models is the way in which ordered
phases break lattice--rotation symmetry.


Given the degree of fine--tuning in the choice of parameters,
it is natural to ask what happens when
$\mathcal{H}^{\sf FFM}_\square$~[Eq.~(\ref{eq:Hex})] is tuned
away from the special line \mbox{$J_3=J_2-|J_1|/2$}
[cf. Fig.~\ref{fig:classical.ground.state.phase.diagram}].
In this case, the collinear and coplanar phases of Fig.~\ref{fig:JTphase} would no longer
belong to the $T=0$ ground--state manifold.
However, commensurate, collinear phases with
\mbox{${\bf Q}=(\frac{2\pi}{3},0)$ or $(0,\frac{2\pi}{3})$} are still generically
more favoured by thermal fluctuations than non--collinear or
non--coplanar phases~\cite{Henley1989}.
And in the case of the Heisenberg antiferromagnet on an anisotropic
triangular lattice, it is known that quantum fluctuations can stabilise a
$1/3$--magnetisation plateau, even where it is not a classical
ground state~\cite{alicea09,coletta13,Chen2013}.


In the present case, our preliminary Monte Carlo simulations hint at a rich
phase diagram near the line \mbox{$J_3=J_2-|J_1|/2$}.
At the lowest temperatures, conical phases with incommensurate wave--vector
predominate.
However at higher temperatures, commensurate, coplanar phases, and in particular
the collinear $1/3$--magnetisation plateau, are restored by fluctuations~\cite{luis-unpub}.
The full determination of the corresponding phase diagram remains an
open problem, and may require use of an algorithm specially adapted to
incommensurability~\cite{Saslow1992}.


Another question of obvious interest is what happens for quantum spins.
For large $S$, it seems reasonable to expect that quantum fluctuations would
act much like thermal fluctuations, stabilising ``triangular--lattice'' physics in
the square--lattice frustrated ferromagnet, both on the special line
$J_3=J_2-|J_1|/2$, and near to it.
The extreme quantum limit of $\mathcal{H}^{\sf FFM}_\square$~[Eq.~(\ref{eq:Hex})], for
spin $S=1/2$, has already been studied in exact diagonalisation~\cite{Sindzingre2009}.
These studies confirm the existence of $1/3$--magnetisation plateau at $T=0$,
for intermediate values of magnetic field, and a range of parameters close to the line
$J_3=J_2-|J_1|/2$.
However, at higher values of field, instead of interpolating to saturation through
a canted phase, the magnetisation plateau gives way to a quantum
spin--nematic phase [cf.~\linecite{Shannon2006,Sindzingre2010}].


It is also reasonable to ask what phases arise for other parameter sets
where the classical ground state of
$\mathcal{H}^{\sf FFM}_\square$~[Eq.~(\ref{eq:Hex})]
is commensurate with the lattice.
These will occur on lines
\begin{eqnarray}
J_1 + 2 J_2  + 4 J_3 \cos \left( \frac{2\pi}{p} \right) = 0
\; ,
\end{eqnarray}
for all integer $p > 3$, with associated ordering vector
\begin{eqnarray}
{\bf Q}_p = \left( \frac{2\pi}{p},0 \right) \; .
\end{eqnarray}
The parameter set ${\bb J}_A$~[Eq.~(\ref{eq:def.JA})], with
three--sublattice order, corresponds to $p=3$.


Our preliminary investigation of this question, using classical Monte Carlo
simulation~\cite{luis-unpub} and exact diagonalisation~\cite{Sindzingre2009},
suggests that the answer is not simple.
For $p=4$, we do indeed find four--sublattice order.
However thermal fluctuations favour a collinear state with ``up--up--down--down''
stripe order over the expected 1D spiral.
This collinear state cants in applied magnetic field, in a way reminiscent
of the planar (XY) model~\cite{Harris1991}.
Quantum fluctuations select the same, collinear, stripe order for
$p=4$ \linecite{Sindzingre2009}.
Meanwhile, exact diagonalisation carried out for
${\bf J} = (-1, 0.3, 0.175)$, i.e.~$p=6$, for clusters of up to $N=36$
spins, indicate that the ground state {\it is} a commensurate 1D spiral.
They also suggest the possibility of a $1/3$--magnetisation plateau
in applied magnetic field.
Clearly, this issue merits further investigation.

\subsection{Spin liquid and ``ring'' correlations}
\label{yyy}

The phase diagram shown in Fig.~\ref{fig:JRphase}, calculated for
the parameter set ${\bb J}_B$, where different spiral phases meet,
looks superficially similar to the one for the parameter set associated with
three--sublattice order ${\bb J}_A$, shown in Fig.~\ref{fig:JTphase}.
However, the addition of a degenerate manifold of
classical ground states leads to a number of crucial differences,
especially at high temperature, and for low values of magnetic field.


The most obvious consequence of the enlarged ground--state manifold
is the ``ring'' seen in the structure factor, as shown in Fig.~\ref{fig:ring-T-evolution}.
This ring is present in both the low--temperature ordered phases, and the
high--temperature paramagnet, for all values of magnetic field.
Simulations for more generic parameter sets~\cite{luis-unpub} confirm
that this ``ring'' is a generic feature of the classical phase boundary between
1D and 2D spiral ground states
[cf. Fig.~\ref{fig:classical.ground.state.phase.diagram}].


At finite temperature, the number of low--lying states contributing to the ring
in the structure factor, $\Sigma_R$ [Eq.~(\ref{SR})], scales with the size of the
system \mbox{[Fig.~\ref{fig:JRingIIA}, \ref{fig:JRing-Nring}]}.
As a consequence, the ring makes a macroscopic contribution to the entropy
of the system, even though the number of classical ground states,
defined by Eq.~(\ref{eq:ring-qx-qy}), only scales with the linear dimension
of the system $L$.
The high--temperature paramagnet phase in Fig.~\ref{fig:JRphase}
can therefore be viewed as a spin liquid with a finite magnetisation.


The spin liquid found in this two--dimensional frustrated ferromagnet
differs from well--known three--dimensional examples, such as spin ice~\cite{henley10},
or the pyrochlore Heisenberg antiferromagnet~\cite{Moessner1998} in that the classical
ground--state manifold is sub--extensive.
It has more in common with problems where ground--state manifold
correspond to lines or surfaces in reciprocal space, such as the chiral spin liquid
found on the diamond lattice~\cite{Bergman2007}, or the ``ring'' and ``pancake''
liquids reported in the frustrated honeycomb antiferromagnet~\cite{Okumura2010}.
We note, however, that unlike the honeycomb lattice example, we do not find
any crossover to a conventional paramagnet at high temperatures.


In the absence of a more complete description, we refer to the new state as a ``helicoidal''
spin liquid, in recognition of the fact that the system displays short--range helicoidal
correlations~\cite{Rastelli1983} --- cf.  Fig.~\ref{fig:JRingIIA}.
The better characterisation of this phase remains an open problem.

\subsection{Conical Spiral}
\label{subsubsection_Conical_Spiral}

Besides supporting a spin liquid, the ground state manifold 
for the parameter set ${\bb J}_B$ [Eq.~(\ref{eq:def.JB})] has the special 
property that it contains a large number of simple, commensurate
wave vectors, ${\bb Q}_{R}$ [Eq.~(\ref{eq:QR})].
This sets the stage for a rather unusual form of order--by--disorder , in which
both single--Q and multiple--Q ordered states are selected by fluctuations
from states belonging to the ``ring'' of allowed ${\bb Q}$ [cf. Fig.~\ref{fig:ring-T-evolution}].
Both of these phases are present for $h=0$, and persist for a finite range 
of magnetic field, as shown in Fig.~\ref{fig:JRphase}.


At the lowest temperatures, thermal fluctuations select an incommensurate 
single--Q state with the character of conical spiral, familiar from studies of 
the Heisenberg antiferromagnet on an anisotropic triangular 
lattice~\cite{Veillette2005,alicea09,Starykh2014,Okubo2012}.
This spiral has uniform magnetisation $S^z$, and wave vector very close 
to ${\bf Q}_{1Q}$ [Eq.~(\ref{eq:Q_1Q})].
In the limit $T \to 0$, we can use a low--temperature expansion
to show that this incommensurate spiral makes a larger contribution 
to entropy than any other state in the ground--state manifold 
of the model (cf. Appendix~\ref{section:selection_1Q_zero_h}).
The same state is found in classical Monte Carlo simulations at low temperatures
$T \lesssim 0.17 J_\Delta$ and low values of magnetic field $h \lesssim J_\Delta$
--- cf. Fig.~\ref{fig:JRphase}.
At low temperatures, but larger values of magnetic field, this incommensurate 
non--coplanar state gives way to the commensurate, coplanar ``Y---state'' familiar 
from the triangular--lattice antiferromagnet 
(cf. Fig.~\ref{fig:JRphase} with Fig.~1 of Ref.~\onlinecite{Seabra2011}).  
 

It is a widely quoted ``rule of thumb'' that fluctuations in frustrated magnets, 
whether quantum or classical, favour coplanar --- and if possible, collinear 
--- states \cite{Henley1989}.
It is also frequently assumed (without proof) that quantum and thermal 
fluctuations will favour the same state.   
The present case provides an interesting counter--example.
Somewhat surprisingly, thermal fluctuations select an incommensurate spiral 
for $T \to 0$, as discussed above.
Meanwhile, as shown in Appendix~\ref{section:selection_quantum_model}, 
quantum fluctuations, treated at the level of linear spin-wave theory, prefer a state 
with commensurate wave vector ${\bb Q}^{\sf 1D}_{\sf 3sub}$ [Eq.~(\ref{eq:def.Q.1D.triangle})]. 


By analogy with the quantum Heisenberg antiferromagnet on a triangular 
lattice \cite{Chubukov_Golosov_1991},  we anticipate that the resulting 
quantum ground state will support three--sublattice, coplanar,  ``120--degree'' 
order for $h = 0$, giving way to a coplanar ``Y---state'' in applied magnetic field.
Thus, for low values of magnetic field, quantum and thermal effects compete, 
setting up the possibility of a phase transition from the commensurate 
quantum ground state into the incommensurate state favoured by thermal
fluctuations, as temperature is increased.
In practice this would mean that the coplanar ``Y---state'', 
already favoured by thermal fluctuations for larger value of field  
--- cf. Fig.~\ref{fig:JRphase} --- would displace the conical spiral for $T \to 0$. 

\subsection{Vortex Crystal}
\label{subsubsection_Vortex_Crystal}

While the conical spiral is a relatively conventional phase, the 
structure of the vortex crystal marks it out as very different from anything 
found in triangular--lattice Heisenberg antiferromagnets.
The vortex crystal is formed through the coherent superposition of classical ground states
with four, symmetry--related wave vectors [Fig.~\ref{fig:JAlowfield}(b)].
It is not selected by harmonic thermal fluctuations
[cf.~Appendix~\ref{section:selection_1Q_finite_h}], but rather by anharmonic (interaction)
effects which become important at higher temperatures.
At a more intuitive level, this $4Q$ state can be visualised as a lattice of
vortices in $S^x-S^y$ plane, with alternating vector chirality [Fig.~\ref{fig:novel.phases}(c)].
The resulting ``crystal'' has a 10--site unit cell [cf.~Fig.~\ref{fig:JAlowfield}(e)],
and meets the usual definition of a magnetic supersolid,
since it spontaneously breaks the translational symmetry of the lattice,
while at the same time breaking spin--rotation symmetry about the direction
of the magnetic field (cf. Refs.~\onlinecite{Seabra2010,Seabra2011}
and references therein).
However it differs from other examples of magnetic supersolids, in that it has a
uniform magnetisation $S^z$, and translational symmetry is instead broken
by the spin texture perpendicular to magnetic field.


Crystals formed of vortices have been widely studied in the context of
type--II superconductors, where they are known
as a ``vortex lattice''~\cite{deGennes-book,Blatter1994}.
Lattices of magnetic vortices are, however, very unusual, recent work on Mott insulators
notwithstanding~\cite{Kamiya2014}.
The $4Q$ state, described above, is not an exact analogue of a superconducting
vortex lattice, since it contains vortices with alternating circulation.
None the less, it is an interesting step in that direction.
The vortex crystal also proves to be a robust state --- we have checked in simulation
that it survives for deviations in exchange parameters $\delta{\bf J} \approx 0.01|J_1|$.
This robustness reflects the fact that it is only possible to form a $4Q$ state
from classical ground states specific to the parameter set ${\bb J}_B$,
ruling out any continuous transformation into an incommensurate state.


Recently, more attention has been paid to a different kind of crystals of topological defects:
lattice states of Skyrmions~\cite{Muhlbauer2009,Schulz2012,Okubo2012,Schutte2014}.
In a Skyrmion lattice the spin texture also also winds in alternating directions, but the core of
each Skyrmion is associated with a modulated $S^z$.
In fact, modulation along the $S^z$  direction seems to be a generic feature of multiple-Q
states found in magnetic models\cite{Okubo2011,Okubo2012}.
This is a strong constraint, and might account for the lack of other examples
of vortex crystals in spin models.
In the light of the recent work on Skyrmion lattices, it might also be interesting
to explore how a vortex crystal couples to itinerant electrons.


It would also be of value to investigate how the $1Q$ and $4Q$ states found
in classical simulations of $\mathcal{H}^{\sf FFM}_\square$
evolve in the presence of quantum fluctuations.
On general grounds, we anticipate that the phases shown in Fig.~\ref{fig:JRphase}
should survive for large (quantum) spin.
In principle, this could be investigated by self consistent mean--field methods.
The picture for spin-$1/2$ is however harder to assess.
Exact diagonalisation studies reveal a very rich ground state phase diagram
in the absence of magnetic field, and suggest a disordered ground state for the
parameter set ${\bb J}_B$ [Refs.~\onlinecite{Sindzingre2009,Sindzingre2010}].
However, given the small cluster sizes available, and the large unit cell of the $4Q$
state, it is impossible to rule out a vortex crystal ground state for spin $S=1/2$.

\section{Conclusions}
\label{section:conclusions}

In this Article we have explored some of the novel phases which arise
in a simple model of a two--dimensional frustrated ferromagnet
--- the $J_1$--$J_2$--$J_3$ Heisenberg model on a square--lattice,
$\mathcal{H}^{\sf FFM}_\square$~[Eq.~(\ref{eq:Hex})].
We have chosen to concentrate on the effect of thermal, rather than quantum
fluctuations, using large--scale classical Monte Carlo simulation, complemented
by linear spin--wave theory, to determine the finite--temperature phase diagram
of the model in applied magnetic field.


Two distinct parameter sets were considered.
The first of these, ${\bb J}_A$~[Eq.~(\ref{eq:def.JA})], was chosen in order
 to favour three-sublattice stripe order.
In this case we find a phase diagram, Fig.~\ref{fig:JTphase}, remarkably
similar to  that of the Heisenberg antiferromagnet on a triangular
lattice~\cite{Seabra2011a}.
In applied magnetic field, a ``120--degree'' ground state transforms
first into a canted ``Y--state'' and then into a collinear $1/3$--magnetisation
plateau, illustrated in Fig.~\ref{fig:novel.phases}(c), before interpolating
through a 2:1 canted phase to saturation.


The second parameter set, ${\bb J}_B$~[Eq.~(\ref{eq:def.JB})], was chosen
to lie at the classical phase boundary between one--dimensional and
two--dimensional spiral ground states.
Here an enlarged ground state manifold manifests itself in a ``ring'' structure
in the spin structure factor, Fig.~\ref{fig:novel.phases}(b), and in a much richer
phase diagram, Fig.~\ref{fig:JRphase}.  
In this case, the phases found include a high--temperature spin--liquid and a number 
of new ordered phases, as well as the ``Y--state'', $1/3$--magnetisation plateau and  
2:1 canted phase familiar from the Heisenberg antiferromagnet on a triangular lattice.
In particular, for low values of magnetic field, a multiple--Q state with the character 
of a vortex crystal [Fig.~\ref{fig:novel.phases}(c)] competes with an incommensurate
conical spiral state.
These results are quite striking, and the subtle play of order by disorder which
drives the finite-temparature phase transition between the incommensurate conical spiral 
and vortex crystal states would have been very difficult to anticipate {\it a priori}.


The abundance of exotic phases found in this simple model
suggests that there is much still to learn about frustrated ferromagnets.
In particular, questions of commensurability, and of the role
of quantum fluctuations, deserve further investigation.
With respect to commensurability, an obvious open question is whether
vortex crystals, like the one studied in this Article, can form
different lattices, or lattices with different lattice constants.
And quantum effects are expected to lead to an even richer
phenomenology, stabilising new forms of magnetic order, as well as 
valence bond solid and spin--nematic states, with exact diagonalisation 
hinting at the possibility of a spin--liquid ground state~\cite{Sindzingre2009}.
And, while the square--lattice Heisenberg model $\mathcal{H}^{\sf FFM}_\square$
provides a simple context for these questions, we anticipate that
the same effects may be observed in a much wider range
of systems.


We conclude with a few comments on experiment.
A number of magnetic insulators have been proposed as examples of
square--lattice frustrated ferromagnets.
These include the vanadium phosphates
Pb$_2$VO(PO$_4$)$_2$~[\onlinecite{Kaul2004}],
BaCdVO(PO$_4$)$_2$~[\onlinecite{Nath2008}],
and SrZnVO(PO$_4$)$_2$~[\onlinecite{Skoulatos2009}],
and the copper--based topotactic ion--exchange systems
(CuCl)LaNb$_2$O$_7$~[\onlinecite{Kageyama2005}],
(CuBr)Sr$_2$Nb$_3$O$_{10}$,~[\onlinecite{Tsujimoto2007}],
and (CuBr)Sr$_2$Nb$_3$O$_{10}$~[\onlinecite{Yusuf2011}].
The properties of the vanadium phosphates in applied magnetic
field appear to well--described by a simple
\mbox{$J_1$--$J_2$} model~\cite{Thalmeier2008,Tsirlin2009}.
%
%
However the much richer phenomenology of the cuprates
suggests more complicated exchange interactions,
which need not have the full symmetry of the
square lattice~\cite{tassel10}.


In the absence of a clear experimental validation from, e.g., inelastic
neutron scattering, the Hamiltonian $\mathcal{H}^{\sf FFM}_\square$~[Eq.~(\ref{eq:Hex})]
is probably best regarded as a toy model capturing some of the
the interesting features of quasi two--dimensional frustrated ferromagnets,
rather than a complete description of the exchange interactions in these
layered cuprates.
None the less, this model does score two notable successes, in providing
a route to both the $1/3$-magnetisation plateau observed
(CuBr)Sr$_2$Nb$_3$O$_{10}$~[\onlinecite{Tsujimoto2007}],
and the helical order found in (CuBr)Sr$_2$Nb$_3$O$_{10}$ [\onlinecite{Yusuf2011}].
In the light of this, it would be very interesting to see further experiments on
square--lattice frustrated ferromagnets in applied magnetic field.

\section*{Acknowledgements}

The authors acknowledge helpful conversations with Hiroshi Kageyama,
Seiji Miyashita, Yukitoshi Motome, Karlo Penc and Keisuke Totsuka.
This work was supported by the Okinawa Institute of Science and Technology
Graduate University (Japan), Funda\c{c}\~{a}o para a Ci\^{e}ncia e Tecnologia (Portugal)
Grant No.~SFRH/BD/27862/2006, Engineering and Physical Sciences Research
Council (United Kingdom) Grants No.~EP/C539974/1 and No.~EP/G031460/1,
Grant-in-Aid for Scientific Research from MEXT (Japan) No.~23540397,
and the RIKEN iTHES Project.
LS, PS and TM gratefully acknowledge the hospitality of the Theory of Quantum
Matter Unit, OIST, where part of this work was completed.


 \input{fic_def.tex}

 \input{appendix_spin_waves.tex}


\end{document}

%% file: fic_def.tex
\newcommand{\rubrique}[1]{\bigskip \noindent {\bf #1}}
\newcommand{\remarque}[1]{{\small \it #1}}

  \newcommand{\bve}[1]{\mathbf{#1}}
  \newcommand{\vve}[1]{\ensuremath{{ \vec{ #1 } }}}

\newcommand{\sousli}[1]{{ \underline{#1} }}
\newcommand{\fichier}[1]{{\em #1}}

\newcommand\arraybslash{\let\\\@arraycr}

\def\ommit#1{}
\def\solution#1{#1}

\newcommand{\ud}{\mathrm{d}}
\newcommand{\udiv}{\mathrm{div}}
\newcommand{\urot}{\vve{\mathrm{rot}}}
\newcommand{\ugrad}{\vve{\mathrm{grad}}}
\newcommand{\prodvec}{\times}
\def\({\left(}\def\){\right)}
\def\cad{c'est-\`a-dire }
\def\bd{\begin{displaymath}}\def\ed{\end{displaymath}}
\def\be{\begin{equation}}\def\ee{\end{equation}}
\def\beq{\begin{equation}}\def\eeq{\end{equation}}
\def\bea{\begin{eqnarray}}\def\eea{\end{eqnarray}}
\def\beqar{\begin{eqnarray}} \def\eeqar{\end{eqnarray}}  
\def\beas{\begin{eqnarray*}}\def\eeas{\end{eqnarray*}}

\newcommand\textsubscript[1]{\ensuremath{{}_{\text{#1}}}}

  \newcommand{\wihat}[1]{{#1}}
  \newcommand{\nohat}[1]{{#1}}
   \def\lan{\langle}
   \def\ran{\rangle}
\newcommand{\ket}[1]{\ensuremath{{| #1 \rangle }}}
\newcommand{\bra}[1]{\ensuremath{{\langle #1 |}}}
\newcommand{\prods}[2]{\ensuremath{{ \langle #1 | #2 \rangle }}}
\newcommand{\elmat}[3]{\ensuremath{{ \langle #1 | #2 | #3 \rangle }}}
\newcommand{\vrho}{\vec{\rho}}
  \newcommand{\vecdot}[1]{\dot{\vve{#1}}}
  \newcommand{\vecddot}[1]{\ddot{\vve{#1}}}
  \newcommand{\vecdddot}[1]{\dddot{\vve{#1}}}
  \newcommand{\nbsublatt}{{K}} 
  \def\yd{^\dagger} 
  \def\nd{^{\vphantom\dagger}}
  \def\sm{{S_{-}}}
  \def\sp{{S_{+}}}
  \def\tS{{\tilde{S}}}
  \def\ij{{\lan ij\ran}} 

\def\d{\delta}
\def\D{\Delta}
\def\t{{\theta}}
\def\a{{\alpha}}
\def\b{{\beta}}

\def\abo{a}  %
\def\vbo{\hat{a}}

\def\va{\bve{ a}}
\def\vr{\bve{ r}}
\def\vk{\bve{ k}}
\def\ve{\bve{ e}}
\def\vmu{\bve{ \mu}}
\def\vm{\bve{ m}}
\def\vex{\bve{ e_x}}
\def\vey{\bve{ e_y}}
\def\vG{\bve{ G}}
\def\vQ{\bve{ Q}}
\def\vq{\bve{ q}}
\def\vS{\bve{ S}}
\def\vM{\bve{ M}}
\def\vR{\bve{ R}}
\def\vRn{{\bve{ R}\nu}}
\def\vxi{\bve{\xi}}
\def\vm{{m}}      
\def\vmn{{m\,\nu}}
\def\cO{{\cal O}}
\def\nup{{\nu'}}
\def\nnp{{\nu\nu'}}
\def\vdelta{{\bve\delta}}
\def\vomega{{\bve\omega}}

\def\kag{{Kagom\'e\ }}
\def\roott{{$\sqrt{3}\times\sqrt{3}$}\ }

\def\cH{{\cal H}}
\def\cA{{\cal A}}
\def\ij{{\lan ij\ran}}
\def\zhat{{\hat z}}
\def\xhat{{\hat x}}
\def\yhat{{\hat y}}
\def\pz{\partial}
\def\etal{{\it et al\/}\ }
\def\ie{{\it i.e.\/}\ }
\def\eg{{\it e.g.\/}\ }
\def\etc{{\it etc.\/}\ }
\def\Neel{N{\'e}el solid\ }
\def\half{\frac{1}{2}}
\def\third{\frac{1}{3}}
\def\twothirds{\frac{2}{3}}
\def\fourth{\frac{1}{4}}
\def\thalf{\frac{3}{2}}
\def\rtot{\frac{\sqrt{3}}{2}}
\def\ta{\theta_{\ssrm A}}
\def\tb{\theta_{\ssrm B}}
\def\tc{\theta_{\ssrm C}}
\def\fa{\phi_{\ssrm A}}
\def\fb{\phi_{\ssrm B}}
\def\fc{\phi_{\ssrm C} }

 \def \sec#1{{Sec.~\ref{#1}}}
 \def \fig#1{{Fig.~\ref{#1}}}
 \def \eqn#1{{Eqn.~(\ref{#1})}}
 \def \tbl#1{{Table~\ref{#1}}}
 \def \app#1{{Appendix~\ref{#1}}}

\def\E{{\rm e}}
\def\I{{\rm i}}

%% file: appendix_spin_waves.tex
%
%

\appendix

\section{Low-temperature entropy in the harmonic approximation and its relation to  spin-wave theory}
\label{section:Appendix_spin_wave}

In this Appendix we provide details of the spin--wave calculations used to explore the 
different ordered phases selected by fluctuations.
We start by reviewing the low--temperature expansion of classical spin model, 
and show how it can be used to explicitly calculate the free energy, 
and thereby the entropy, associated with a given form of magnetic order, 
to leading order in $T$ [Section~\ref{subsection:Entropy_harmonic_approximation}].  


We then show how equivalent results for the free energy can be obtained from the 
spin--wave dispersion obtained within linear spin--wave theory for the associated 
quantum model, using the conical spiral state as a worked example 
[Section~\ref{subsection:LSW}].  


Finally, we derive explicit expressions for the entropy associated with 
the different forms of magnetic order encountered in this work~:
(i)~conical spirals in applied magnetic field [Section~\ref{spin_waves_Conical_spirals}],
(ii)~coplanar $\nbsublatt$-sublattice states with in--plane magnetic field, 
such as the $Y$ state [Section~\ref{Coplanar_K_sublattices_states}], 
(iii)~canted conical $\nbsublatt$-sublattice states
such as the 4Q state in applied field [Section~\ref{Canted_K_sublattices_states}].

\subsection{Entropy within a classical low--temperature expansion}
\label{subsection:Entropy_harmonic_approximation}

The low-temperature thermodynamics of a classical spin model of the form
Eq.~(\ref{eq:Hex})
can be calculated through an expansion of the  Hamiltonian
in small fluctuations about the ground state ($T=0$) configuration.
For this purpose, one may proceed as follows.
Let us consider a magnetic ground state  with $\nbsublatt$ sublattices.
Without loss of generality, it is possible to select a reference frame
such that, in this ground state,
the spin on the $\nu^{th}$ sublattice ($\nu=1,\cdots,\nbsublatt$) in the $m$th unit cell
at position $\vR_{\nu,m}$ is written as
\beq
\label{ground_state_conf}
\vS^{0}_{\nu,m}=
\begin{bmatrix}
 S^x_{\nu,m} \\
 S^y_{\nu,m} \\
 S^z_{\nu,m}
\end{bmatrix} =
\begin{bmatrix}
\sin(\theta_{\nu,m}) \sin (\phi_{\nu,m} ) \\
\cos(\theta_{\nu,m}) \\
\sin(\theta_{\nu,m}) \cos (\phi_{\nu,m} )
\end{bmatrix}
\;.
\eeq
We then choose for each site $(\nu,m)$
a local frame $(\bve{\tilde{e}}_x,\bve{\tilde{e}}_y, \bve{\tilde{e}}_z)$,
rotated from the reference frame,
such that $\bve{\tilde{e}}_z$ coincides with the
ground-state spin direction $\vS^{0}_{\nu,m}$,
and expand the spin components in this frame as
\beq
\vS_{\nu,m}=
\begin{bmatrix}
 x_{\nu,m} \\
 y_{\nu,m} \\
  z_{\nu,m} \approx 1
   - \frac{ x^2_{\nu,m} + y^2_{\nu,m} }{2}
\end{bmatrix}  \; ,
\label{eq:classical_expansion}
\eeq
where $x_{\nu,m}$ and $y_{\nu,m}$ denote transverse components of a spin deviation.
One has, in the harmonic approximation,
\beq
\label{eq:H_hamo}
 \mathcal{H}=  \mathcal{H}^{(0)} + \mathcal{H}^{(2)},
\eeq
where $\mathcal{H}^{(0)}$ is the energy  of the ground state
and $\mathcal{H}^{(2)}$ is quadratic in spin deviations.
For $N$ spins with $N_c$ unit cells~($N_c=N/\nbsublatt$):
\beq
\label{eq:H_quadra}
 \mathcal{H}^{(2)}  = \frac{1}{2} \hat{\vxi}^T \mathcal{M} \hat{\vxi},
\eeq
where $\hat{\vxi}$
is a $2N$-dimensional vector of spin deviations,
\beq
\hat{\vxi}^T = [x_{1,1},\cdots,
x_{\nbsublatt,Nc} ,
           y_{1,1},\cdots ,
           y_{\nbsublatt,Nc}]
\eeq
and $\mathcal{M}$ a symmetric $2N\times 2N$ matrix.
By Fourier transforming the spin deviations
\beq
   x(y)_{\nu,m}   = \frac{1}{\sqrt{N_c}}
   \sum_{\vk} x(y)_{\nu,\vk}  \E^{i \vk \cdot \vR_{\nu,m}} ,
\eeq
Eq.~(\ref{eq:H_quadra})  becomes
\bea
 \mathcal{H}^{(2)} &=&
  \frac{1}{2}
  \sum_{\vk\in M_{\rm BZ}}    \hat{\vxi}^T_{-\vk}  \mathcal{M}(\vk)  \hat{\vxi}_{\vk}
\label{eq:Csw-H}
\eea
where
\beq
\hat{\vxi}_{\vk}^T= [x_{1,\vk}, ,\cdots,x_{K,\vk},y_{1,\vk},\cdots, y_{K,\vk}].
\eeq
Here $M_{\rm BZ}$ denotes the magnetic Brillouin zone for the \mbox{$K$-sublattice} structure,
and
$\mathcal{M}(\vk)$ is a $2\nbsublatt\times 2\nbsublatt$ Hermitian matrix.
Diagonalizing $\mathcal{M}(\vk)$ by a unitary transformation, one obtains
the form
\beq
\label{eq:Csw-H_diag}
 \mathcal{H}^{(2)} = \frac{1}{2}\sum^{2\nbsublatt}_{\nu}
\sum_{\vk \in M_{\rm BZ} }\kappa_{\nu,\vk} \zeta_{\nu,\vk} \zeta_{\nu,-\vk},
\eeq
where
$\kappa_{\nu,\vk}$ are the eigenvalues of $\mathcal{M}(\vk)$.
The free energy per site at low $T$ is written as
\beq
\label{eq:free-energy}
\frac{\mathcal{F}}{N}=\frac{E_0}{N} -T\ln T
- T \frac{S_{sw}}{N}
+ {\mathcal O}(T^2) \; ,
\eeq
where 
\begin{align}
 \frac{S_{sw}}{N} =
 - \lan \ln \kappa_{\nu,\vk}\ran =
 -  \frac{1}{2N} \sum^{2\nbsublatt}_{\nu=1}   \sum_{\vk}
    \ln \kappa_{\nu,\vk} \;,
\label{eq:def_S_sw}
\end{align}
is the entropy per spin associated with harmonic spin fluctuations\cite{footnote-3}.   
We note that the evaluation of  $\lan\ln \kappa_{\nu,\vk}\ran$
does not require the computation of the eigenvalues of $\mathcal{M}(\vk)$
but only of its determinant $|\mathcal{M}(\vk)|$, since
 \begin{eqnarray}
 \lan \ln \kappa_{\nu,\vk}\ran &=&
 \frac{1}{2N} \sum_{\vk} \ln
 \left[\prod_{\nu}^{2\nbsublatt} \kappa_{\nu,\vk}  \right]
\nonumber\\
 &=& \frac{1}{2N} \sum_{\vk} \ln |\mathcal{M}(\vk)|
   \;.
\label{eq:def_kappa_av}
\end{eqnarray}

\subsection{Alternative derivation of entropy from linear spin--wave theory
for a quantum model}
\label{subsection:LSW}

It is also possible to evaluate the determinant $|\mathcal{M}(\vk)|$
[Eq.~(\ref{eq:def_kappa_av})], which determines the low--temperature entropy
of classical spins [Eq.~(\ref{eq:def_S_sw})], starting from a linear spin--wave
(LSW) theory for quantum spins.
To show how this works, we first derive the large--$S$, LSW expansion of
$\mathcal{H}^{\sf FFM}_\square$ [cf. Eq~(\ref{eq:Hex})], and then take
a classical limit, setting $S=1$.


We consider the same ground state configuration with $K$ sublattices as discussed in
Section~\ref{subsection:Entropy_harmonic_approximation}.
After the rotation to the local frames and a Holstein-Primakoff transformation
of spin operators into bosonic creation and annihilation operators
$a_{\nu,\vk}^\dagger$ and $a_{\nu,\vk}$,
\begin{align}
\vS_{\nu,m}\approx
\begin{bmatrix}
 \sqrt{S/2}(a_{\nu,m}+a_{\nu,m}^\dagger) \\
 -i\sqrt{S/2}(a_{\nu,m}-a_{\nu,m}^\dagger) \\
 S - a_{\nu,m}^\dagger a_{\nu,m}
\end{bmatrix}  \; ,
\label{eq:HP_transform}
\end{align}
the harmonic Hamiltonian for spin $S$
is written in the form
\begin{equation}
\label{eq:H_sw}
 \mathcal{H}^{(2)}_{\rm qu} =  \frac{1}{2} \sum_{\vk} \Bigl[
\hat{\bf a}_{\vk}^\dagger M({\vk}) \hat{\bf a}_{\vk}-\Delta_{\vk} \Bigr]
\;,
\end{equation}
where
\beq
 \hat{\bf a}_{\vk}^\dagger =
[a_{1,\vk}^\dagger\ldots ,a_{\nbsublatt,\vk}^\dagger,
a_{1,-\vk},\ldots ,a_{\nbsublatt,-\vk}],
\eeq
$M(\vk)$ denotes a $2K\times2K$ matrix,
and $\Delta_{\vk}$ is a scalar function
which determines the zero--point energy associated with 
a given form of order.


We can diagonalise $\mathcal{H}^{(2)}_{\rm qu}$ [Eq.~(\ref{eq:H_sw})] 
using a (para-unitary) Bogoliubov transformation\cite{Colpa}
\beq
\label{eq:Bogoliubov_1}
\hat{\bf a}_{\vk} =T_{\vk} \hat{\bf b}_{\vk}\;,
\eeq
such that
\beq
\label{eq:para_Id_1}
 T^\dagger_{\vk} I_{-1}  T_{\vk}=I_{-1},
\eeq
with
\beq
\label{eq:mat_Im1}
 I_{-1} =
 \begin{bmatrix}
          I_K &  0 \\
           0 & -I_K
  \end{bmatrix},
\eeq
and a $K\times K$ identity matrix $I_K$,
brings
 $\mathcal{H}^{(2)}$ into a diagonal form
\beq
\label{eq:H_sw_diag}
 \mathcal{H}^{(2)}_{\rm qu}=   \frac{1}{2} \sum_{\vk} \Bigl[
\hat{\bf b}_{\vk}^\dagger \Omega({\vk}) \hat{\bf b}_{\vk}
 -\Delta_{\vk} \Bigr],
\;
\eeq
with
\beq
 \Omega({\vk})= I_{-1}T^{-1}_{\vk} I_{-1} M(\vk) T_{\vk} =
 \begin{bmatrix}
          \omega_{\vk} &  0 \\
           0 & \omega_{-\vk}
  \end{bmatrix},
\label{eq:mat_Omega}
\eeq
where  $\vomega_{\vk}$
is the  diagonal matrix
\beq
\label{eq:omega}
 \vomega_{\vk} =
 \begin{bmatrix}
 \omega_{1,\vk} &            &        &  0 \\
            & \omega_{2,\vk} &        &  \\
            &                & \ddots &  \\
          0 &                &        & \omega_{\nbsublatt,\vk}
 \end{bmatrix}
\;
\eeq
and $\omega_{\nu,\vk}$ is the spin wave frequency in the $\nu$
branch at wave vector $\vk$.


We now consider a classical limit in Eq.~(\ref{eq:H_sw}) by setting $S=1$, in order to
compare it with the classical harmonic Hamiltonian Eq.~(\ref{eq:H_quadra}).
In the classical limit the spin operators $a_{\nu,\vk}$  and $a_{\nu,\vk}^\dagger$ are
replaced with complex conjugate scalar fields $\psi_{\nu,\vk}$  and $\psi_{\nu,-\vk}^\ast$.
(This corresponds to considering a path integral formulation and omitting the imaginary-time dependence
in the fields.)
In this classical limit, the Holstein-Primakoff transformation Eq.~(\ref{eq:HP_transform})
of $S=1$ spins in the lowest order
becomes equivalent to the expansion Eq.~(\ref{eq:classical_expansion}) of classical spins,
if we set
\begin{eqnarray}
\psi_{\nu,\vk} &=& ( x_{\nu,\vk} + i y_{\nu,\vk}) /\sqrt{2} \; , \nonumber\\
\psi_{\nu,-\vk}^\ast &=& ( x_{\nu,-\vk} - i y_{\nu,-\vk}) /\sqrt{2} \; .
\end{eqnarray}


Thus the classical limit of the quantum $S=1$ harmonic Hamiltonian
becomes equivalent to the
classical harmonic Hamiltonian under the transformation
\beq
 \hat{\bf \psi}_{\vk} = \mathcal{T} \hat{\vxi}_{\vk},
\eeq
with
\beq
\hat{\bf \psi}_{\vk}^T= [\psi_{1,\vk}^\ast,\cdots,\psi_{K,\vk}^\ast, \psi_{1,\vk},\cdots, \psi_{K,\vk}],
\eeq
and
\beq
\label{def:mathcal_T_1}
  \mathcal{T} = \frac{1}{\sqrt{2}} \begin{bmatrix}
        1 & &        & 0 & -i & &        & 0 \\
          & 1 &      &   &    & -i &     &   \\
          & & \ddots &   &    & & \ddots &   \\
        0 & &        & 1 &  0 & &        & -i\\
        1 & &        & 0 &  i & &        & 0 \\
          & 1 &      &   &    & i &      &   \\
          & & \ddots &   &    & & \ddots &   \\
        0 & &        & 1 &  0 & &        & i \\
  \end{bmatrix}
\;.
\eeq
Since
$\Delta_{\vk}$ originates from commutation relations of operators, it
 can be neglected in the classical limit. Comparing both classical Hamiltonians, one finds that
\beq
  \hat{\vxi}^T_{-\vk}  \mathcal{M}(\vk)  \hat{\vxi}_{\vk} =
  \hat{\bf \psi}_{-\vk}^T M({\vk}) \hat{\bf \psi}_{\vk},
\eeq
which entails
\beq
\label{relation:M_mathcal_M}
  \mathcal{M}({\vk}) = \mathcal{T}^\dagger M(\vk) \mathcal{T}.
\eeq
From this relation one immediately obtains
\beq
\label{relation:det_M_det_mathcal_M}
|\mathcal{M}({\vk})|= |M({\vk})|,
\eeq
since $|\mathcal{T}^\dagger| |\mathcal{T}| =1$, and from
Eq.~(\ref{eq:mat_Omega})
\beq
\label{relation:det_M_det_Omega}
|M({\vk})| = |\Omega(\vk)| = \prod_{\nu}^{\nbsublatt} \omega_{\nu,\vk} \omega_{\nu,-\vk}.
\eeq
Setting $S=1$, we obtain the relation
\beq
|\mathcal{M}({\vk})| =
\prod_{\nu}^{\nbsublatt} \omega_{\nu,\vk} \omega_{\nu,-\vk}, \;
\label{eq:det_M_K_sublattice}
\eeq
With this result in place, the entropy $S_{sw}$ entering into the classical 
free energy $\mathcal{F}$ [Eq.~(\ref{eq:free-energy})], can be 
calculated from the linear, quantum spin--wave dispersion 
$\omega_{\nu,\vk}$ [Eq.~(\ref{eq:omega})], through 
\begin{eqnarray}
\frac{S_{sw}}{N}
	&=& - \lan\ln \kappa_{\vk}\ran
		= \frac{1}{2N} \sum_{\bf k} \ln |\mathcal{M}({\vk})| \nonumber\\
	&=& -\frac{1}{N} \sum_{\nu=1}^{\nbsublatt}  \sum_{\vk} \ln \omega_{\nu,\vk}
		= -\lan\ln \omega_{\nu\vk}\ran \; .
\label{eq:kappa_av_K_sublattice}
\end{eqnarray}
We will explicitly show in the next subsection
that Eq.~(\ref{eq:det_M_K_sublattice}) holds for spiral states.
We note that the spin-wave frequencies $\omega_{\nu,\vk}$ 
could also be obtained by solving the semi--classical equations
of motion for the quantum spin--1 model.


In conclusion, the classical entropy per spin, $S_{sw}/N$, can be calculated 
from either the determinant of the matrix $\mathcal{M}(\vk)$ [Eq.~(\ref{eq:def_kappa_av})] 
found within a classical low--temperature expansion, or the values of 
the dispersion $\omega_{\nu,\vk}$ obtained within LSW 
theory [Eq.~(\ref{eq:omega})].
The first approach is slightly simpler but the latter
is also interesting as it
provides  information on the dynamics both for the classical
and the quantum model in the semi-classical approximation.
It may also be more rapidly implemented if the LSW theory
of the model has been derived previously.

\subsection{Conical spirals}
\label{spin_waves_Conical_spirals}

For the case of  conical spirals, it is convenient
to choose the reference frame such that
the  magnetic field $\bve{h}$ is along the $S^y$ axis, instead of
the $S^z$ axis.
The ground-state spins of the conical spirals  have then the same
projection on the $S^y$ axis,
whereas their projections on the $S^x$-$S^z$ plane describe a spiral with
wave vector $\vQ$.
So, they can  be written as in Eq.~(\ref{ground_state_conf}), with
equal $\theta$  for all sites,
introducing just one sublattice ($\nbsublatt=1$) as
$\phi_{1,m} = \phi_m = {\vQ} \cdot {\vR_m} $, where $\vR_m$ is the
location of the site $m$.
After the rotation to the local frames
$(\bve{\tilde{e}}_x,\bve{\tilde{e}}_y, \bve{\tilde{e}}_z)$
and the expansion in spin deviations,
the harmonic Hamiltonian reads
\begin{align}
 \mathcal{H}^{(2)} & = \frac{1}{2}
  \sum_{\vk}    \hat{\vxi}^T_{-\vk}  \mathcal{M}(\vk)  \hat{\vxi}_{\vk}
 \nonumber\\
\label{eq:H_quadra_cone_2}
 &= \frac{1}{2}  \sum_{\vk}
   [x_{-\vk},y_{-\vk}]
 \begin{bmatrix}
 \mathcal{M}^{xx}(\vk) & \mathcal{M}^{xy}(\vk) \\
 \mathcal{M}^{yx}(\vk) & \mathcal{M}^{yy}(\vk) \\
 \end{bmatrix}
 \begin{bmatrix}
 x_{\vk} \\ y_{\vk}
 \end{bmatrix}.
\end{align}
Here $\mathcal{M}(\vk) $ is the $2 \times 2$ Hermitian matrix with coefficients
\begin{align}
 \mathcal{M}^{xx}(\vk) &= \frac{1}{2}[J(\vk+\vQ) + J(\vk-\vQ)] - J(\vQ),
 \nonumber\\
 \mathcal{M}^{yy}(\vk) &= \mathcal{M}^{xx}(\vk)  \cos^2\theta
          + \left[J(\vk) - J(\vQ) \right] \sin^2\theta,
 \nonumber\\
  \mathcal{M}^{xy}(\vk) &=  (M^{yx}(\vk))^{\ast} =
   \frac{1}{2i}\left[ J(\vk+\vQ) - J(\vk-\vQ) \right] \cos\theta,
\label{coef_mat_M_cone}
\end{align}
where  $J(\vk)$ is the Fourier transform of the interactions,
given in Eq.~(\ref{eq:FT.interactions}).
The canting angle $\theta$ is fixed by
\beq
 \cos\theta = \frac{h}{J(0) -J(\vQ)}  \;.
\eeq

%

Next we describe the LSW theory of the quantum \mbox{spin--$S$} model.
After the rotation to the local frames and a Holstein-Primakoff transformation
of spin operators into bosonic creation and destruction operators
$a_{\vk}^\dagger$ and $a_{\vk}$, the harmonic Hamiltonian for spin $S$
is written as~\cite{coletta13,Coletta2013}
\begin{equation}
 \mathcal{H}^{(2)}_{\rm qu} =  \frac{1}{2} \sum_{\vk} \Bigl[
\hat{\bf a}_{\vk}^\dagger M({\vk}) \hat{\bf a}_{\vk}-\Delta_{\vk} \Bigr]
\;,
\end{equation}
where
\beq
\hat{\bf a}_{\vk}^\dagger = (a_{\vk}^\dagger,a_{-\vk})
\eeq
and
\beq
\Delta_{\vk}=-SJ(\vQ)
\eeq
[cf. \linecite{note_app_01}].
The $2\times2$ matrix $M(\vk)$ is given by
\begin{equation}\label{eq:mat_M}
 M({\vk})=
 \begin{bmatrix}
          A({\vk})+C({\vk}) &  B({\vk}) \\
           B({\vk}) & A({\vk}) -C({\vk})
  \end{bmatrix},
\end{equation}
with
\bea\label{eq:Coeff_M_helix}
\nonumber
A({\vk}) &=&
  \frac{S}{4} \big\{(\cos^2\theta+1) [J(\vk+\vQ)+J(\vk -\vQ)]
   -4 J(\vQ)   \\ \nonumber
   & &+
  2\sin^2\theta J(\vk)\big\}, \\
\nonumber
B({\vk}) &=&
  \frac{S}{4}\sin^2\theta[ J(\vk+\vQ)+J(\vk -\vQ) - 2 J(\vk) ],\\
C({\vk}) &=&
    \frac{S}{2}\cos\theta[J(\vk+\vQ) -J(\vk -\vQ)]
\;.
\eea
A  Bogoliubov transformation
\beq
\hat{\bf a}_{\vk} =T_{\vk} \hat{\bf b}_{\vk},\;
\eeq
brings
 $\mathcal{H}^{(2)}_{\rm qu}$ into a diagonal form
\beq
\label{eq:H_sw_diag_cone}
 \mathcal{H}^{(2)}_{\rm qu}=   \frac{1}{2} \sum_{\vk} \Bigl[
\hat{\bf b}_{\vk}^\dagger \Omega({\vk}) \hat{\bf b}_{\vk}
 -\Delta_{\vk} \Bigr]
,\;
\eeq
with
\beq \label{eq:mat_Omega_cone}
 \Omega({\vk})=
 \begin{bmatrix}
          \omega_{\vk} &  0 \\
           0 & \omega_{-\vk}
  \end{bmatrix},
\eeq
where the spin wave dispersion is
\beq
\label{eq:omega_k}
\omega_{\vk} = \sqrt{[A(\vk)+B(\vk)][A(\vk)-B(\vk)]} + C(\vk).
\eeq


In the case of $S=1$ we have the following relations
\bea
\mathcal{M}^{xx}({\vk}) &=& A({\vk})+B({\vk}),
\nonumber\\
\mathcal{M}^{yy}({\vk}) &=& A({\vk})-B({\vk}),
\nonumber\\
\mathcal{M}^{xy}({\vk}) &=&  i C({\vk}) \;.
\label{relation:mathcal_M_ABC}
\eea
One can indeed directly see
from Eqs.~(\ref{eq:omega_k}) and (\ref{relation:mathcal_M_ABC})
that
\bea
 \omega_{\vk} \omega_{-\vk} &=& A^2({\vk})-B^2({\vk}) - C^2({\vk})
\nonumber\\
&=&   \mathcal{M}^{xx}({\vk}) \mathcal{M}^{yy}({\vk})
 -|\mathcal{M}^{xy}({\vk})|^2
 \nonumber\\
&=&  |\mathcal{M}({\vk})|,
\;
\eea
and hence
\begin{eqnarray}
\frac{S_{sw}}{N}
	&=& - \frac{1}{N} \sum_{\vk} \ln \omega_{\vk}
	= - \lan \ln \omega_{\vk} \ran \;.
\label{eq:kappa_av_one_sublattice}
\end{eqnarray}

\subsection{Spin wave dispersion in coplanar $\nbsublatt$-sublattice states}
\label{Coplanar_K_sublattices_states}

We now turn to the case of coplanar ground states such as the $Y$ state
with an in-plane  field
(or the coplanar $4Q$ state in zero field)
which require the introduction of $\nbsublatt$ sublattices.
For the sake of greater generality we rewrite  the Hamiltonian
as
\beq
\mathcal{H} =
\sum_{\lan\nu,m;\nu',m'\ran}  J_{\nu,m;\nu',m'}
\vS_{\nu,m}\cdot\vS_{\nu',m'}
- \bve{h}\cdot\sum_{\nu,m} \vS_{\nu,m} \;.
\eeq
The Fourier transform of the exchange couplings is defined by
 \beq
 \label{eq_def_J_nunup}
 J_{\nu,\nu'}(\vk) = \sum_{m} J_{\nu,0;\nu',m }
    \E^{ i \vk \cdot (\vR_{\nu',m} -  \vR_{\nu,0})}.
 \eeq
Choosing the reference frame such that the ground state spins lie
in  the $S^x-S^z$ plane,  with $\bve{h}$ along the $S^z$ axis,
the ground-state spins  can then be written as in
Eq.~(\ref{ground_state_conf}),  with $\theta=\pi/2$ for all sites
and $\phi_{\nu,m}=\phi_{\nu}$ independent of the unit cell.
After expressing the spin deviations in the local frames,
one obtains a harmonic Hamiltonian of the form Eq.~(\ref{eq:H_quadra})
with
\beq
\label{mat_M_K_sublattices}
\mathcal{M} (\vk) =
\begin{bmatrix}
 \mathcal{M}^{xx}(\vk) & \mathcal{M}^{xy}(\vk) \\
 \mathcal{M}^{yx} (\vk)& \mathcal{M}^{yy}(\vk)\\
 \end{bmatrix},
\eeq
where  $\mathcal{M}^{\alpha\beta}(\vk)$ ($\alpha,\beta=x,y$)
are $\nbsublatt\times\nbsublatt$ matrices  given by
\bea
\label{coef_mat_M_K_sublattices}
\nonumber
 \mathcal{M}^{xx}_{\nu,\nu'}  (\vk) &=& J_{\nu,\nu'}(\vk) \cos\phi_{\nu,\nu'}
 +\delta_{\nu,\nu'}
 \left[ h\cos\phi_{\nu} - N_{\nu} \right],
 \\ \nonumber
 \mathcal{M}^{yy}_{\nu,\nu'} (\vk)  &=& J_{\nu,\nu'} (\vk)
 +\delta_{\nu,\nu'}
 \left[ h\cos\phi_{\nu} - N_{\nu} \right],
 \\ %
 \mathcal{M}^{yy}_{\nu,\nu'} (\vk)  &=& \mathcal{M}^{yx} (\vk) = 0,
\eea
with $\phi_{\nu,\nu'} = \phi_{\nu} - \phi_{\nu'}$ and
\begin{align}
 N_{\nu} &=   \sum_{\nu'} J_{\nu,\nu'}(0) \cos\phi_{\nu,\nu'}.
\label{eq_diag_N}
\end{align}


There are $\nbsublatt=3$ sublattices in the $Y$ state.
The expressions for the angles $\phi_{\nu}$  are similar to those  found
for the triangular antiferromagnet
\begin{align}
\label{eq:angles_Y_state}
 \phi_{1} &=\pi,\nonumber\\
 \phi_{2} &=-\phi_{3}=\arccos \frac{1}{2}\left( \frac{h}{3J_{\triangle}} +1 \right),
\end{align}
 where
$J_{\triangle}$ is given in
Eq.~(\ref{eq:def.J.triangle}),
whereas the exchange couplings
 Eq.~(\ref{eq_def_J_nunup})  are
\bea
\label{eq:coef_J_nunup_Y_state}
   \nonumber
 J_{\nu,\nu}(\vk) &=& J_1\cos(k_y) + J_3\cos(2k_y)\,, \quad \{\nu=1,2,3\}
  \\ \nonumber
 J_{1,2}(\vk) &=&  J_{2,3}(\vk)  = J_{3,1}(\vk)
  \\
 &=& \frac{J_1}{2}\E^{i k_x} + J_2 \E^{i k_x } \cos(k_y)
   + \frac{J_3}{2}\E^{-2 i k_x } \;.
\eea
For the $4Q$ states the number of sublattices is $\nbsublatt=10$ and
we do not show the matrix of couplings $J_{\nu,\nu'}(\vk)$.

\subsection{Spin wave dispersion in canted $\nbsublatt$-sublattice states}
\label{Canted_K_sublattices_states}

Here we consider canted conical states with $\nbsublatt$ sub lattices.
As for the case of conical spirals, the reference frame is chosen such that the
magnetic field
is along the $S^y$ axis and the ground-state spins
can then be written as in Eq.~(\ref{ground_state_conf}). These canted
conical states
have  equal projections of the spins along the field
 (equal values of $\theta$  for all sites)
whereas the spin projections in the $S^x$-$S^z$ plane form a
$\nbsublatt$ sublattice configuration
and include, as a special case, the $4Q$ state  where the spins
uniformly cant in the direction of the magnetic field.


The ground state has equal $\theta$  for all sites
and $\phi_{\nu,m}=\phi_{\nu}$ independent of the unit cell.
The matrix $\mathcal{M} (\vk)$ has the same structure  as
in Eq.~(\ref{mat_M_K_sublattices}) but with
\bea
\label{coef_mat_M_K_sublattices_canted}
 \nonumber
\mathcal{M}^{xx}_{\nu,\nu'}  (\vk) &=& J_{\nu,\nu'}(\vk) \cos\phi_{\nu,\nu'}
 -\delta_{\nu,\nu'} N_{\nu},
 \\ \nonumber
 \mathcal{M}^{yy}_{\nu,\nu'} (\vk)  &=& \mathcal{M}^{xx}_{\nu,\nu'}(\vk)   \cos^2\theta
+  \left[ J_{\nu,\nu'}(\vk) - \delta_{\nu,\nu'} N_{\nu} \right] \sin^2\theta,
 \\ %
 \mathcal{M}^{xy}_{\nu,\nu'} (\vk)  &=&
\left[ J_{\nu,\nu'}(\vk) - J_{\nu,\nu'}(-\vk) \right] \sin\phi_{\nu,\nu'} \cos\theta,
\eea
where the canting angle $\theta$ is related to the applied field by
\beq
 \label{ef_J_nunup}
 \cos\theta  = \frac{h \nbsublatt }{
   \sum_{\nu\nu'}   J_{\nu,\nu'}(0) (1-\cos\phi_{\nu,\nu'} )
 }
\;.
\eeq
In the case of a commensurate
spiral state, $\phi_{\nu} =\vQ\cdot\vR_{\nu,0}$, one can check that, as expected, 
$\mathcal{H}^{(2)}$ [Eq.~(\ref{eq:H_quadra})] 
can be rewritten in the form Eq.~(\ref{eq:H_quadra_cone_2})
with the coefficients  Eq.~(\ref{coef_mat_M_cone})
where
\beq
 J(\vk) =
 \sum_{\nu}  J_{1,\nu}(\vk) \E^{ i \vk \cdot \vR_{\nu,0}  }
\;.
\eeq
For the same reasons as  above,  Eq.~(\ref{eq:kappa_av_K_sublattice}) holds
and the entropy factor can be alternatively computed from the spin-wave
frequencies.

\subsection{Spin--wave dispersion in spiral states}
\label{subsection:One-sublattice dispersion}

In the main text we investigate the fingerprints of the spin waves of spiral states in the uniform paramagnetic phase.
For this purpose, the spin-wave spectrum %
of the fully polarized state provides useful information about the low-energy excitations above the ring.
The dispersion relation is obtained at the saturation field in a straightforward manner  as
\begin{align}
\omega( {\bf k} )_{\sf 1SL}
 &= J({\bf k}) - J({\bf Q}) ,
 \label{eq:dispersion1SL}
\end{align}
which has the lowest energy at the spiral wave vector ${\bf Q}$.
By imposing a low-energy cutoff, %
we can identify momenta close to the ring,
which contribute to the low-energy excitations, and  use them in Eq.~(\ref{SR}) to calculate
the total structure factor associated with the ring.


\begin{figure}[t]
\centering
\includegraphics[width=\columnwidth]{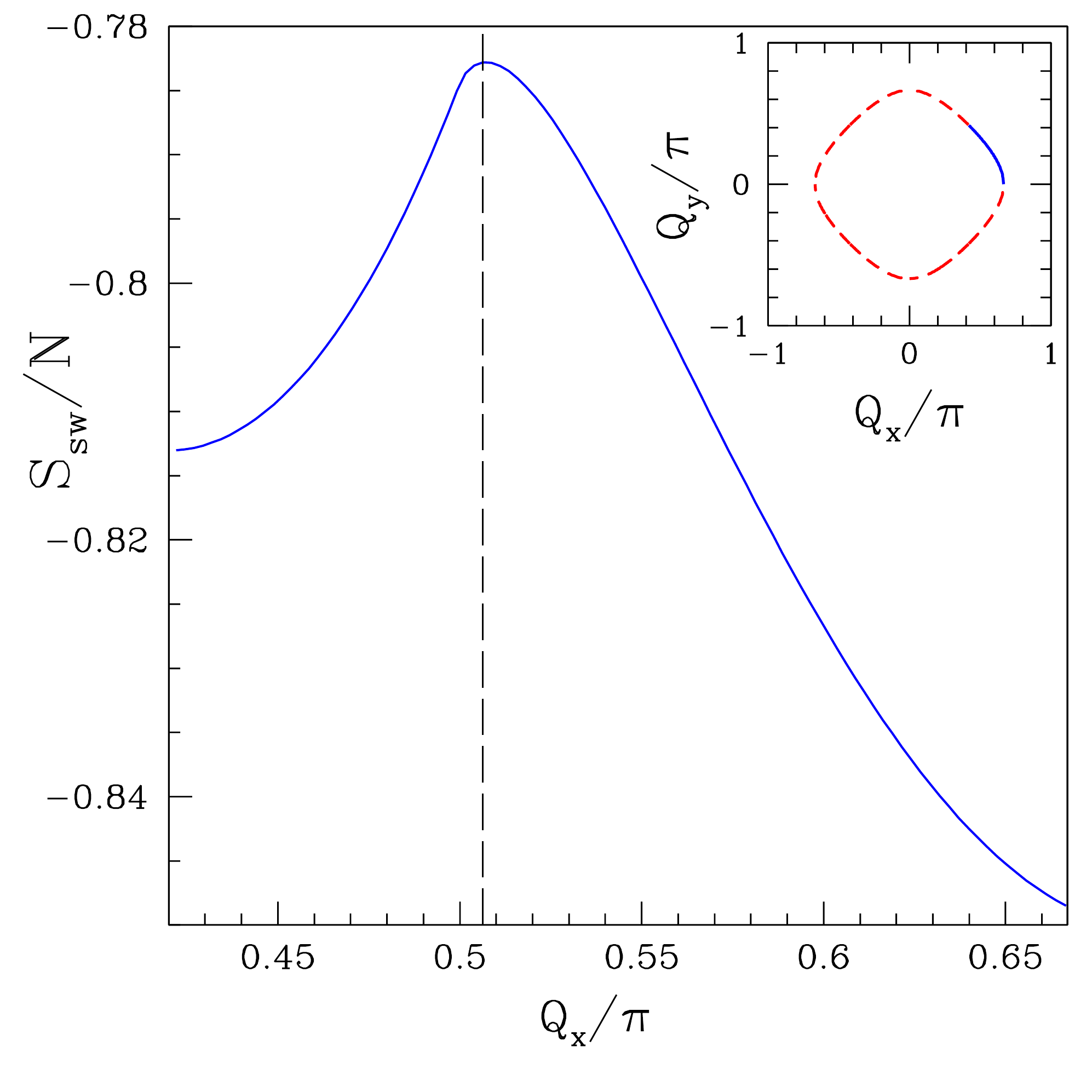}
\caption{
\footnotesize (Color online)
Entropy per spin $S_{sw}/N$ of the two--dimensional spiral states which make up
the classical ground state manifold of $\mathcal{H}^{\sf FFM}_\square$
[cf. Eq~(\ref{eq:Hex})] for the parameter set ${\bf J}_B$ [Eq.~(\ref{eq:def.JB})]
The inset shows the set of wave vectors associated with the ground--state
manifold, as defined by Eq.~(\ref{eq:ring-qx-qy}).
Thermal fluctuations select the state with highest entropy, which corresponds
to a an incommensurate spiral with wave vector close to  $(\pi/2,\pi/3)$.
$S_{sw}/N$ was calculated through Eq.~(\ref{eq:kappa_av_one_sublattice}),
using the linear spin--wave dispersion $\omega (\vk)$ [Eq.~(\ref{eq:omega_k})],
as discussed in Appendix~\ref{section:Appendix_spin_wave}.  
\label{fig:CSW-I}
}
\end{figure}


\begin{figure}[t]
\centering
\includegraphics[width=\columnwidth]{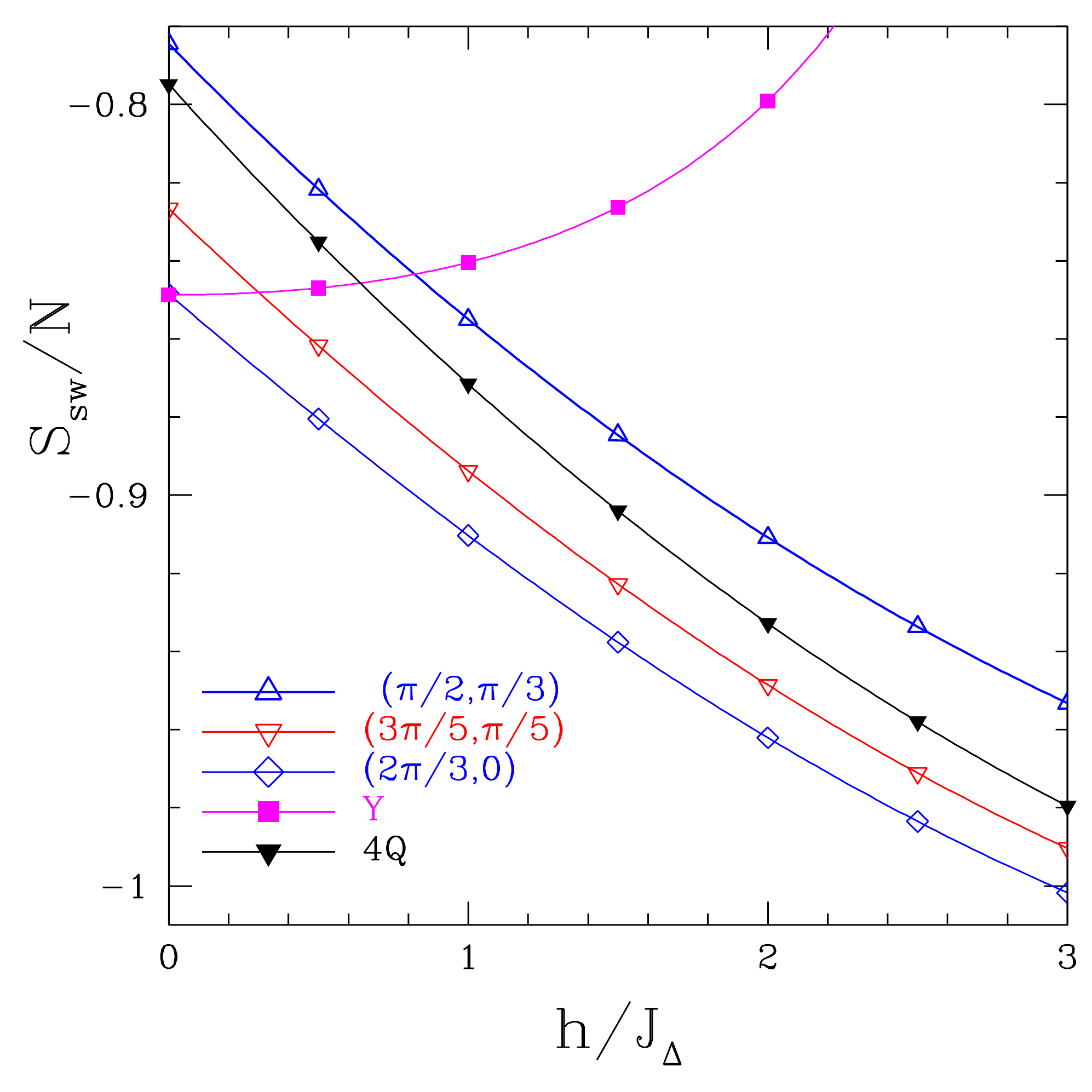}
\caption{ \footnotesize(Color online)
Entropy per spin $S_{sw}/N$ [Eq.~(\ref{eq:kappa_av_one_sublattice})]
of different competing states as a function of magnetic field $h$.
The entropy of the coplanar $Y$ state increases with increasing field
and,  for \mbox{$0.8 J_\triangle < h < 3 J_\triangle$}, overtakes that of
the conical state with \mbox{${\bf Q} = (\pi/2, \pi/3)$}.
Entropy was calculated with linear spin--wave (LSW) theory,
for the parameter set ${\bf J}_B$ [Eq.~(\ref{eq:def.JB})], 
as discussed in Appendix~\ref{section:Appendix_spin_wave}.  
Within this linear, low--temperature approximation, the 4Q state is
not selected by fluctuations at any value of magnetic field.
\label{fig:CSW-II}
}
\end{figure}


It would be interesting to see how this dispersion is modified in the spiral ordered phase.
Here we consider the coplanar spiral state at zero field. The dispersion is obtained from
Eq.~(\ref{eq:omega_k}) by setting $S=1$ and $\theta=\pi/2$
\begin{align}
\omega( {\bf k} )
 &= \sqrt{ \frac{1}{2} [J(\vk+\vQ) + J(\vk-\vQ) - 2 J(\vQ) ]
 \left[J(\vk) - J(\vQ) \right]
 }
.
\end{align}
When ${\bf k}$ is close to the spiral wave vector ${\bf Q}$, one finds
\begin{align}
\omega( {\bf k} )
 &\approx \sqrt{\textsf{J}[J({\bf k}) - J({\bf Q})]},
\end{align}
where $\textsf{J} = [J(2\vQ) + J(0)]/2 - J(\vQ) $.
At the parameter set ${\bf J}_{B}$, the spin-wave dispersion has gapless line modes
along ${\bf k}={\bb Q}^{\sf ring}$ [Eq.~(\ref{eq:ring-qx-qy})] in the LSW approximation.

\section{Entropy selection of a spiral with $\vQ\approx(\pi/2,\pi/3)$ out of the 
degenerate ring in the harmonic approximation}
\label{section:selection_1Q_zero_h}

In this Appendix we show how, for the parameter set ${\bb J}_B$~[Eq.~(\ref{eq:def.JB})], 
the harmonic analysis developed in Appendix \ref{section:Appendix_spin_wave} predicts 
the entropic selection of a spiral state at low temperatures and vanishing magnetic field, 
as suggested by Monte Carlo simulations.
In Fig.~\ref{fig:CSW-I} we plot the the entropy per spin $S_{sw}/N$ [Eq.~(\ref{eq:def_S_sw})]
for the the family of conical states with wave-vector $\vQ=(Q_x,Q_y)$ satisfying
the ring equation Eq.~(\ref{eq:ring-qx-qy}), in the absence of an applied 
magnetic field. 
This was computed from the spin-wave frequencies [Eq.~(\ref{eq:kappa_av_one_sublattice}) 
and Eq.~(\ref{eq:omega_k})],  and is $S_{sw}/N$ plotted as a function of $Q_x$.


Thermal fluctuations select,  within the degenerate ring,
the spiral state with the highest entropy, which is described by
an incommensurate wave-vector slightly away from
$(\pi/2,\pi/3)$, or wave-vectors related to it by symmetry.
This result is supported by the Monte-Carlo simulations
[see Sec.~(\ref{subsection:low-field-phase-diag-jt})], and 
continues to hold for a small applied field, where the state evolves
from a coplanar into a conical spiral.

\section{Entropy competiton between the  $\vQ\approx(\pi/2,\pi/3)$ 1Q cone
and the coplanar $Y$ states in the harmonic approximation}
\label{section:selection_1Q_finite_h}

In this Appendix we show how, for the parameter set 
${\bb J}_B$~[Eq.~(\ref{eq:def.JB})], a canted $Y$ state 
prevails over a conical version of the spiral state at low temperatures, for a sufficiently 
strong applied magnetic field.
In Fig.~\ref{fig:CSW-II}, we plot the entropy per spin $S_{sw}/N$  [Eq.~(\ref{eq:def_S_sw})]
for a selection of different states within the degenerate ground state manifold, as a 
function of magnetic field, calculated using the results of Appendix~\ref{spin_waves_Conical_spirals}
and Appendix~\ref{Canted_K_sublattices_states}.   
The  entropy factor of the coplanar $Y$ state increases with increasing field,
until it is finally selected by at low-temperature fluctuations for
$0.8J_\triangle\lesssim h<3J_\triangle$ as supported by the
Monte-Carlo simulations [see Sec.~(\ref{subsection:low-field-phase-diag-jt})].
The 4Q state is not favoured by thermal fluctuations at low temperature, which
is again in agreement with the simulation results.
However, its entropy factor at low field is rather close to the one
of the most favored conical spirals.
This small difference is overcome by
anharmonic effects which stabilize the 4Q  state at larger $T$
for  field  $h \lesssim 0.8J_\triangle$.

\section{Quantum selection out of the degenerate ring in the  quantum  model in the large--$S$ limit.}
\label{section:selection_quantum_model}

In this Appendix we discusses the role of quantum fluctuations in selecting 
an ordered ground state from the degenerate ground state manifold for the 
parameter set ${\bb J}_B$~[Eq.~(\ref{eq:def.JB})], within the linear spin--wave
(LSW) theory developed in Appendix \ref{subsection:LSW}.
From the LSW expansion of the quantum model
 [Eq. (\ref{eq:H_sw_diag})], one finds that
the ground-state energy per spin in the large-$S$ limit
can be written as
\beq
\label{eq:E_qu_general}
 e_{\rm qu} = e_{0} + \Delta e,
\eeq
where $e_{0}$ is the ground-state energy per spin of the classical model
and $\Delta e$ is its first-order correction in $1/S$ due to the
the zero point motion of the spin waves
\beq
\label{eq:delta_e_qu}
\Delta e = \frac{1}{2N} \sum_{\vk} \Bigl[{\rm Tr}   (\vomega_{\vk}) - \Delta_{\vk}\Bigr].
\eeq
Here $\Delta e$ is proportional to $S$ whereas $e_{0}$ is proportional to $S^2$.
Among a set of classically degenerate ground state, quantum fluctuations
will favor the state which minimizes $\Delta e$.


\begin{figure}[tb]
\centering
\includegraphics[width=\columnwidth]{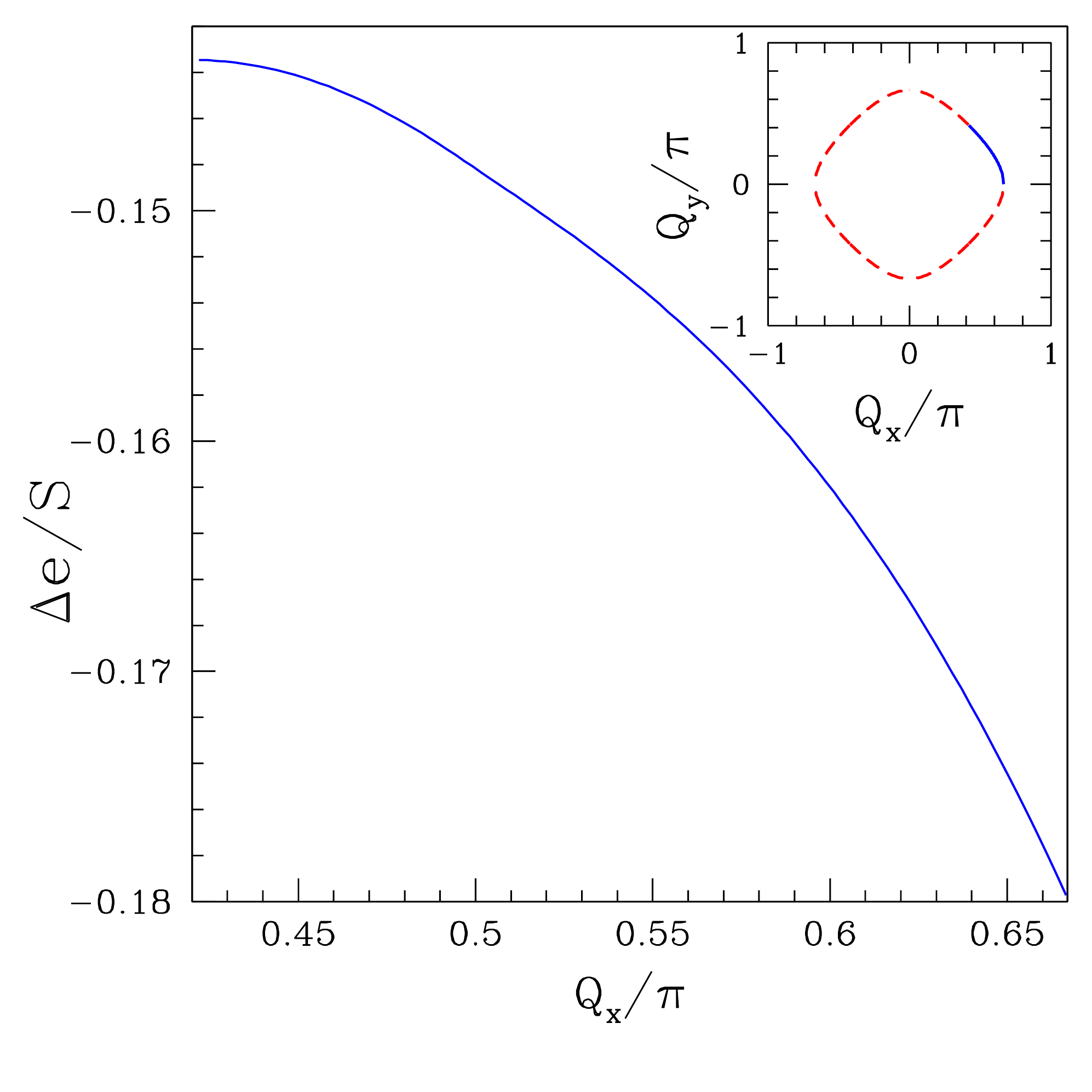}
\caption{
\footnotesize (Color online)
Quantum correction $\Delta e/S$
to the ground-state energy (per spin) of spiral states
[Eq.~(\ref{eq:delta_e_conical_spirals})]
for the spin-$S$ quantum model discussed in Appendix \ref{subsection:LSW}. 
Calculations were carried out within the linear spin--wave theory (LSW), 
for coplanar sprial states with wave vector $\vQ$ belonging to the classical 
ground--state manifold in the absence of magnetic field, for the parameter 
set ${\bf J}_B$ [Eq.~(\ref{eq:def.JB})].
The inset shows the range of wave vectors considered.
Quantum fluctuations select a sprial with $\vQ=(2\pi/3,0)$, 
and symmetry--related states.
\label{fig:CSW-III}
}
\end{figure}


For conical spirals
\beq
\label{eq:e0_conical spirals}
e_{0} =  \frac{S^2}{2} \left[ J(0) \cos^2\theta  + J(\vQ)\sin^2\theta \right]
        + h S \cos\theta,
\eeq
which reduces to $e_{0} = S^2 J(\vQ) /2$ in zero field,
and
\beq
\label{eq:delta_e_conical_spirals}
 \Delta e = \frac{1}{2N} \sum_{\vk} \omega_{\vk}  + \frac{S}{2}  J(\vQ).
\eeq
%
In Fig.~\ref{fig:CSW-III} we show $\Delta e/S$ at the ${\bf J}_{B}$ point in zero field
for the family of conical states with wave-vector $\vQ=(Q_x,Q_y)$
satisfying
the ring equation Eq.~(\ref{eq:ring-qx-qy}).
One sees that quantum fluctuations favor the three-sublattice
stripe states with  $\vQ=(\pm 2\pi/3,0)$ or $\vQ=(0,\pm 2\pi/3)$.

A similar calculation based on the LSW theory
for the $\nbsublatt$-sublattice cases [Eq.~(\ref{eq:H_sw_diag})]
shows that,  in zero field, the $4\vQ$ state has a higher energy
than the $\vQ=(2\pi/3,0)$ spiral state
(or equivalently the $Y$ state at zero field).
In a  finite field, the $Y$ state  becomes favored
over the conical spirals and the canted  $4\vQ$ state.


%% file: paper.bbl
\begin{thebibliography}{99}
\bibitem{footnote-3} The entropy per site is  $\frac{S_{sw}}{N} + \ln T$
which is the average of the logarithm of a dimensionless ratio.
\bibitem{Coletta2013}
Our expression of $B({\vk})$ in Eq.~(\ref{eq:Coeff_M_helix}) has opposite sign
 as  we choose the  spiral  to cant toward  the $y$  axis.
\end{thebibliography}
